# Automatic detection of multilevel communities: scalable and resolution-limit-free

Kun Gao, Xuezao Ren, Lei Zhou and Junfang Zhu

*School of Science, Southwest University of Science and Technology, Mianyang 621010, Sichuan Province, People's Republic of China*

Corresponding authors:   Kun Gao (Email: kgao@mail.ustc.edu.cn)

Xuezao Ren (Email: rxz63@aliyun.com)

**Abstract**

Community structure is one of the most important features of complex networks. Modularity-based methods for community detection typically rely on heuristic algorithms to optimize a specific community quality function. Such methods are limited by two major defects: (1) the resolution limit problem, which prohibits communities of heterogeneous sizes being simultaneously detected, and (2) divergent outputs of the heuristic algorithm, which make it difficult to differentiate relevant and irrelevant results. In this paper, we propose an improved method for community detection based on a scalable community "fitness function." We introduced a new parameter to enhance its scalability, and a strict strategy to filter the outputs. Due to the scalability, on the one hand our method is free of the resolution limit problem and performs excellently on large heterogeneous networks, while on the other hand it is capable of detecting more levels of communities than previous methods in deep hierarchical networks. Moreover, our strict strategy automatically removes redundant and irrelevant results, without any artificial selection. As a result, our method neatly outputs only the stable and unique communities, which are largely interpretable by the *a priori* knowledge about the network, including the implanted structures within synthetic networks, or metadata for real-world networks.

**Keywords:** community detection, resolution limit problem, modularity, multilevel community, Louvain algorithm

# 1. Introduction

Community, also known as network cluster, is a mesoscopic structure ubiquitous in many real-world systems whose topologies are generally described by complex networks [1]. Since highlighted by Girvan and Newman in 2002 [2], community structure of network has been of particular interest to physicists and mathematicians, as it characteristically reveals functional, relational or even social information of complex systems [1, 3, 4, 5]. Despite of certain special definitions of community (for instance, the "disassortative structures" as studied in [6]), communities are most typically defined as groups of nodes with the connections inside each group being denser than those between different groups [7]. Community detection explores optimized divisions of a network. In previous literature, it has become a standard practice to evaluate the effectiveness of a community detection method by its performances in either recreating the implanted communities in synthetic (artificial) networks, or recovering observed node attributes or metadata for real-world networks [8].

To detect the "correct" communities for a network is always challenging. Related studies can be traced back all along to "graph partitioning" in graph theory [9], or "hierarchical clustering" in sociology [10]. For large graphs, finding an exact solution to a partitioning task has been proven an NP-complete problem [7, 9]. And in the case of real-world complex networks, it is even harder: the total number of communities is usually unknown [7], the sizes of different communities may differ by orders of magnitude [11], and the overall structure of the whole network is often multilevel or hierarchical [11, 12, 13]. Community detection methods generally turn to heuristic algorithms for acceptably good solutions [7]. Existing methods can be broadly categorized into either probabilistic ones or non-probabilistic ones [14]; among them, the most popular methods include stochastic block models [15, 16], modularity-based methods [7, 17, 18, 19, 20], and information-based approaches such as Infomap [21]. Within the scope of this paper, we only study modularity-based methods. In a broad sense, modularity-based methods assess the validity of each potential network division with a specific community quality function. This quality function, is in practice formulated into either a modularity function or fitness function [7, 17, 18, 19, 20], or a Hamiltonian of a first principle Potts model [22, 23, 24, 25, 26, 27]. Heuristic algorithm is then employed to optimize the quality function, i.e., to maximize the modularity or fitness function, or minimize the Potts Hamiltonian.

Modularity-based methods, including the Potts models, have been argued to be limited by two major defects. The first one is the well-known "resolution limit problem" raised by Fortunato and Barthelemy in 2007 [28]. Depending on the numbers of intra connections of communities and the total number of connections within the whole network, a modularity optimization method tends to merge small communities (even if they are well-defined clusters as complete graphs) into larger but sparser ones. This reveals a fact that modularity-based method can't find communities of small sizes, like a microscope can't find microbes beyond its resolution range. For quality functions other than the modularity, the same phenomenon has also been observed [29]. On the other hand, modularity-based methods may detect unreasonable community structures due to inappropriate resolutions, for example they detect communities for random graphs [30, 31]. In order to overcome these problems, a variety of "multiresolution" methods [3, 20, 24, 26] and "resolution-limit-free" methods [27, 32] have been suggested. Multiresolution methods use a tunable parameter to alter the resolution; they detect communities of different levels in different resolution scales. Resolution-limit-free methods refrain from using a null model, which can also relieve the resolution limit [33]. However, a further study by Lancichinetti and Fortunato [33] pointed out that the resolution limit problem is actually induced by two opposite tendencies: the tendency of merging small communities, and the tendency of breaking large ones. When communities in a same network have very different sizes, it becomes impossible for a quality optimization approach to avoid both biases simultaneously. Multiresolution and resolution-limit-free methods seemed to outperform other methods only because the community sizes used in their tests were "too close to one another," spanning less than one order of magnitude [33]. When the community sizes vary over up to two orders of magnitude, as in many real-world networks [11, 34], existing multiresolution and resolution-limit-free methods also fail to detect the expected community structures [33].

The second defect of modularity optimization methods is that finding an optimal division for any given network is normally infeasible. It has been recognized that the modularity landscape of a network often includes an exponentially growing (with system size) numbers of local maxima [33, 35]. These local maxima may all be very close to the global maximum in terms of modularity, but the corresponding divisions of the network can be topologically utterly different from one another [35]. This implies that not only an exactly optimal division for a network is intrinsically unreachable [33], but the available solutions in practice can be largely unstable and inconsistent. Multiresolution approaches furtherly aggravate the problem: communities detected in "irrelevant" resolution scales are often "messy:" they are mostly incomprehensible and, for all practical purposes, uninformative. Although it has been argued that inquiring which is the "best" or most "relevant" scale of resolution is an ill posed question [3], many methods still manage to find out the most *stable* communities that can be detected within a persistent range of resolution. The existence of such stable communities is an observed fact [3]: the numbers of communities form strong "plateaus" along the resolution scales [26]. It has become popular in previous literature to rank the relevance of detection results by their strengths of plateaus; community structures suggested by strong plateaus have been demonstrated to be frequently consistent with the *a priori* knowledge about the network [3, 20, 26].

Nevertheless, existing methods using the stability of plateaus are not yet satisfying. On the one hand, a "plateau" is supposed to reflect multiple times of convergence onto an *identical* topology of community structure at *different* resolutions. However, no explicit comparison on the topologies of different data points within a same plateau has been presented in previous literature, leaving a doubt that these data points may possibly represent not the same communities at all. Some methods as in [3, 36, 37] simply distinguish plateaus by the numbers of communities, some by the values of the modularity [20], while some others as in [26] execute information-based quantitative comparisons among the topologies of communities detected on multiple "replicas" of the network at each *fixed* resolution—yet communities detected at varying resolutions are still not compared. According to the discussion in [35], none of such definitions can guarantee each plateau as defined represents an identical community structure. On the other hand, evaluating the stability of plateaus by their lengths is not always effective. Although large plateaus are almost all stable [3, 20, 26], stabilities of small plateaus are uncertain: some of them can be stable and informative, but some others just emerge due to randomness (we will show some examples in Result and Discussion). Previous literature simply ignores all small plateaus, stable or unstable, or artificially selects their preferred results to interpret. Due to the above facts, we believe a strict definition for the term "plateau," as well as an effective strategy to evaluate the stabilities of plateaus, are both urgently needed.

In this paper, we propose our new approach for multiresolution community detection. We adopt a modified community fitness function [20] and a heuristic Louvain algorithm [38] to find multilevel community structures for complex networks. We suggest a strict strategy to identify the "best-and-unique solutions" automatically, and organize them into stable plateaus. As a result, our approach is scalable and resolution-limit-free, with neat outputs. It performs well on both synthetic benchmark networks and real-world networks.

## 2. Method

Our method is a modularity-based multiresolution method, which includes three components: (1) a modified community fitness function with a tunable resolution parameter and a scaling factor, (2) a heuristic Louvain algorithm to maximize this community fitness function, and (3) a strategy to filter the output and retrieve the most stable and significant results. Next we introduce these three components separately.

### 2.1 Community fitness function

The so-called community fitness function was firstly proposed by Lancichinetti and Fortunato [20]. Its original form is as following:

$$f^{\mathcal{G}} = \frac{k_{in}^{\mathcal{G}}}{(k_{in}^{\mathcal{G}} + k_{out}^{\mathcal{G}})^{\alpha}}.$$  (Formula 1)

Here $\mathcal{G}$ denotes a community given by a certain network division, $f^{\mathcal{G}}$ quantifies the fitness (i.e., quality) of community $\mathcal{G}$: larger values of $f^{\mathcal{G}}$ indicate more reliable communities. $k_{in}^{\mathcal{G}}$ and $k_{out}^{\mathcal{G}}$ in the formula stand for the *in-degree* and *out-degree* of community $\mathcal{G}$, defined in the same way as those in previous literature such as [13]. $\alpha$ ($\alpha>0$) is a *resolution parameter* that tunes the resolution: large values of $\alpha$ yield small communities, while small values of $\alpha$ deliver large communities [20].

Instead of the widely-used modularity $Q$ proposed by Newman [7, 39], we choose to use this community fitness function because it is by design scalable and is promising to avoid the resolution limit problem. The original form as shown in formula 1 can be directly used in our method; actually we do use it in many of our calculations in the following part of this paper. Yet for complex networks having multilevel or hierarchical structures, it is indeed helpful to introduce an additional parameter to rescale the varying range of resolution parameter $\alpha$. In this paper, we adopt a modified form of $f^{\mathcal{G}}$ as following:

$$\left(f_{\alpha}^{\beta}\right)^{\mathcal{G}} = \frac{(k_{in}^{\mathcal{G}})^{\beta}}{(k_{in}^{\mathcal{G}} + k_{out}^{\mathcal{G}})^{\alpha}}.$$  (Formula 2)

Here the power exponent $\beta$ ($\beta \geq 1$) is our newly introduced "*scaling factor*;" when $\beta=1$, formula 2 degrades to formula 1. When detecting communities with $\left(f_{\alpha}^{\beta}\right)^{\mathcal{G}}$, we search for an optimized division of the given network that maximizes the *summative fitness* of all detected communities:

$$F_{\alpha}^{\beta} = \sum_{\mathcal{G}} \left(f_{\alpha}^{\beta}\right)^{\mathcal{G}} = \sum_{\mathcal{G}} \frac{(k_{in}^{\mathcal{G}})^{\beta}}{(k_{in}^{\mathcal{G}} + k_{out}^{\mathcal{G}})^{\alpha}}.$$  (Formula 3)

Alternatively, to maximize the *average fitness* ($F_{\alpha}^{\beta}$ divided by the total number of communities) as in [20] also yields essentially equivalent results.

The scaling factor $\beta$ amplifies the varying range of the resolution parameter $\alpha$. For each fixed $\beta$, we estimate in the supplementary material that the "relevant" range of $\alpha$ should be $\beta$-1<$\alpha$<2$\beta$-1. Here the upper bound 2$\beta$-1 prevents inappropriate splitting of large communities—one important example is, random graphs exhibit no community structure with $\alpha$<2$\beta$-1. And the lower bound $\beta$-1 avoids unexpected merging of small communities: dense clusters that are sparsely connected will not be combined into a large community. Obviously, when $\beta=1$, the relevant range of $\alpha$ is between 0 and 1; but when $\beta>1$, the corresponding relevant range of $\alpha$ has been amplified $\beta$ times.

In practice, for networks having only one single community level, varying $\alpha$ within ($\beta$-1, 2$\beta$-1) is already sufficient for the expected communities being detected. But for networks with multiple community levels, within the relevant range ($\beta$-1, 2$\beta$-1) only communities of the lowest level (i.e., communities of smallest sizes that cannot split further) can be detected. This is because higher levels of communities come from combinations of lower-level communities; to detect such combinations, the lower bound $\beta$-1 must be relaxed. Therefore, in our calculation we run our algorithm with the resolution parameter $\alpha$ varying between 0 and 2$\beta$-1, which enables us to detect multiple levels of communities within different resolution scales.

**2.2 Heuristic optimization algorithm**

Since optimizing a community quality function has been proven an NP-complete problem [7, 9], heuristic algorithms are generally adopted to obtain the *best available solutions*. Early methods such as those in [2, 7] usually have heavy demands on computational resources, while more recently a number of faster algorithms have been proposed [17, 34, 38, 39, 40]. Among them, the "Louvain" algorithm raised by Blondel *et al.* [38] is widely accepted due to its prominent efficiency and high accuracy. The label "Louvain" comes from the authors' affiliation (UCLouvain); alternatively, it is also called a "BGLL" algorithm by the authors' initials. Originally, this algorithm is designed as a greedy algorithm to optimize the standard modularity function *Q* proposed by Newman [39]; similar algorithms have also been adopted to optimize the Potts Hamiltonian by other methods [26, 27].

We employ the Louvain algorithm to optimize our community fitness function (formula 3) in this paper. Here we briefly describe its steps; for more details, please refer to [38].

(1) *Initialize communities*. At the very beginning, each node of the network is designated to an individual community. A network consisting of *N* nodes is then divided into *N* communities of size 1.

(2) *Optimize communities of the lowest level*. Sequentially consider each node of the network and scan its neighboring communities (i.e., communities sharing at least one edge with the node in focus). Calculate the potential gains of $F_\alpha^\beta$ if the node in focus was moved out of its original community and put into each of the neighboring communities. Place the node in focus into the community that leads to a maximum value of $F_\alpha^\beta$.

(3) *Iterate until convergence*. Repeat step (2) until a maximum value of formula 3 is reached where no more moves of any node may further increase this value. During this process, the sequence of node orders is randomized every time a new round of iteration is started.

(4) *Merge communities to build a higher community level*. Consider each community obtained at the convergence of step (3) as a fixed module; hereafter all its members (nodes) must be moved together. Repeat the above steps (2) and (3) by taking each fixed module as a node. During this process, connected modules gradually condense into communities of higher levels, until a maximum value of formula 3 is reached.

(5) *Iterate until convergence at the highest level*. Repeat step (4) and detect communities of all levels, until the highest level is detected where no further merging of any communities can increase $F_\alpha^\beta$.

(6) *Output communities*. Communities of all levels detected by the above steps (1) to (5) form a hierarchical structure; each level can be independently outputted. Customarily, only the output of the highest level is adopted since it has a maximum value of $F_\alpha^\beta$ among all levels.

**2.3 Strategy to filter the output**

As a heuristic approach, the Louvain algorithm optimizing our community fitness function may converge to different solutions in different realizations. Many of these solutions are "local maxima," which emerge only by chance. To retrieve from "messy" outputs the most "relevant" solutions that can ben persistently detected, previous methods customarily study the stability of "plateaus" [3, 20, 26, 35, 36, 37]. Yet the term "plateau" was so far *everywhere loosely defined* (see our arguments in the Introduction). Thus we suggest a much stricter definition for "plateau," and correspondingly a strict strategy to identify them. By our definition, a plateau is a continuous scale of resolution within which a heuristic optimization algorithm uniformly converges to a *unique* topological structure of community. To identify such plateaus, it is required to compare the topologies of not only solutions obtained at each fixed resolution, but also those obtained at different resolutions. We suggest the following strategy to discover our plateaus:

(1) *At each fixed resolution* (i.e., with fixed values of parameters *α* and *β*), we implement the Louvain algorithm on the same network in multiple realizations. Among the outputs of all realizations, we adopt the ones with the highest value of $F_\alpha^\beta$ as our *best solutions* obtained at this resolution. In addition, we require the topology of these best solutions must be *unique*: in case two or more solutions have equally highest values of $F_\alpha^\beta$, but represent even slightly different topological structure of community, all solutions obtained at this resolution will be abandoned, and the corresponding resolution will be considered "irrelevant" and does not contribute to any potential plateau.

(2) *At different resolutions*, with varying values of *α* (during which the value of *β* is still fixed), we run the above step (1) and obtain the *best-and-unique* solutions at all *relevant* resolutions. Then we compare the topologies of these best-and-unique solutions, and classify them into different plateaus: solutions in a same plateau must represent exactly the same topological structure of community.

The above step (1) compares the topologies of communities detected at fixed resolutions. We require the uniqueness of our best solutions because non-unique solutions often result from *random convergences* of the Louvain algorithm; below we will show some examples. Step (2) compares the topologies of the best-and-unique solutions obtained at varying resolutions. Plateaus defined and identified as above strictly fulfil "*one plateau, one topological structure of community;*" our strict strategy guarantees such plateaus are truly stable, since random convergences of the algorithm have all been abandoned. Only then can the stability (or say, "robustness") of solutions be measured by the lengths of their corresponding plateaus. It is common in previous literature [3, 20, 26, 35, 36, 37] to either take the communities detected by the longest plateaus as the most relevant results, or artificially select favored structures that best fit the *a priori* knowledge about the network. To our viewpoint, all best-and-unique solutions represented by plateaus should have their own particular information [3]: they discover different structures and aspects of the given network. Thus in this paper, we do not select plateaus simply by their lengths. Instead, we output and interpret "only the stable plateaus, and all the stable plateaus."

## 3. Results

In this section, we firstly demonstrate the effectiveness of our method in recovering implanted community structures for synthetic benchmark networks, including multilevel communities within hierarchical networks (section 3.1), and communities of distinct sizes in heterogeneous networks (section 3.2). Then we apply our method to real-world networks, exhibiting its consistency with the observed node attributes or metadata (section 3.3).

**3.1 On the hierarchical Ravasz-Barabasi (RB) networks**

We first of all introduce the structure of the Ravasz-Barabasi (RB) networks [11]. The smallest RB network is RB5, which is a complete graph consisting of 5 nodes and 10 edges, see figure 1 (a). To facilitate our discussion, we call node 0 the *central node*, while all other nodes *peripheral nodes* of the RB5 network. RB5 is a basic unit to constitute larger RB networks. Five RB5 units, one in the center and four on the periphery, constitute an RB25 network, as shown in figure 1 (b) and (c). According to [11], these RB5 units should be connected in such a way: every peripheral node of the peripheral RB5 units is connected to the central node of the central RB5 unit, but the peripheral RB5 units themselves are not connected to one another. Note that in figure 1 (b) and (c), for easy drawing we did not shift the central node of each RB5 unit slightly off its center as in figure 1 (a): each RB5 unit in figure 1 (b) and (c) is still a complete graph containing 10 edges, only the diagonal edges are invisible by overlapping with other edges. Following the same way, five RB25 units constitute an RB125 network, as shown in figure 2. Obviously, the RB networks are fractal-like and hierarchical, which can grow infinitely. We choose these networks to test our method because they can provide hierarchical structures of any depth—deep enough to test the limit of any multiresolution method for community detection.

As synthetic/artificial networks, the implanted community structures within the RB networks are apparent: an RB network with $5^n$ nodes (hereafter we call it an "RB$5^n$ network") is expected to be naturally divided into $5^{n-1}$ RB5 units on the lowest community level, or $5^{n-2}$ RB25 units on a higher level, and so on, constructing a hierarchical structure of $n$ levels. However, such "natural" divisions should not be taken for granted. One problem is, within each RB$5^m$ unit ($2 \leq m \leq n$), the central node (e.g., see the hollow red circles in figures 1 (c), 2 (b), (c) and (d)) is connected to *every* peripheral node of the same unit. By the natural division, this central node always has a larger out-degree than in-degree (for example, the central node of an RB25 network as in figure 1 (b) has an out-degree of 16 but an in-degree of 10), which violates the customary definition of community in a strong sense [41]. Moreover, within larger networks, this problem gets even more serious: when $n \geq 3$ the natural division further violates the definition of community in a weak sense [41]. Figure 2 (a) shows an example: the central RB5 community (solid red circles) within an RB125 network has a total out-degree of 80 (all contributed by its central node), but a total in-degree of only 20. Therefore, it seems reasonable to break up the central RB5 community and segregate its central node as an individual community, as we did in figure 1 (c), which has also been suggested previously by [42].

Through literature investigation (such as on [42]), as well as numerical simulation, for the RB networks, we propose a plausible revision to their "natural divisions," i.e., on each community level, preserve the peripheral communities, but further divide the central communities. More specifically, on the lowest community level, instead of 5 communities (as in figure 1 (b)), we divide each RB25 unit into 6 communities (as in figure 1 (c)). Thus an RB$5^n$ network ($n \geq 2$) will be divided into $6 \times 5^{n-2}$ communities (since it contains $5^{n-2}$ RB25 units); figure 2 (b) shows an example that an RB125 network is so divided into 30 communities. Similarly, on a higher level, we find each RB125 unit tends to split into 10 communities as in figure 2 (c): 4 peripheral RB25 units make 4 communities, while the central RB25 unit splits to 6 communities as in figure 1 (c). On this level, an RB$5^n$ network ($n \geq 3$) will be divided into $10 \times 5^{n-3}$ communities. Following this regulation, on the $m$-th community level ($2 \leq m \leq n$, here $m=1$ stands for the lowest level), it is each RB$5^{m+1}$ unit that splits into $4m+2$ communities, thus an RB$5^n$ network ($n \geq m+1$) will be divided into $(4m+2) \times 5^{n-m-1}$ communities. Supplementary figure 1 (a) visualizes an RB625 network being divided into 14 communities on the third community level, and table 1 summarizes the numbers of communities on different levels of the RB networks given by the above divisions. Next we demonstrate how such divisions would be discovered by our method.

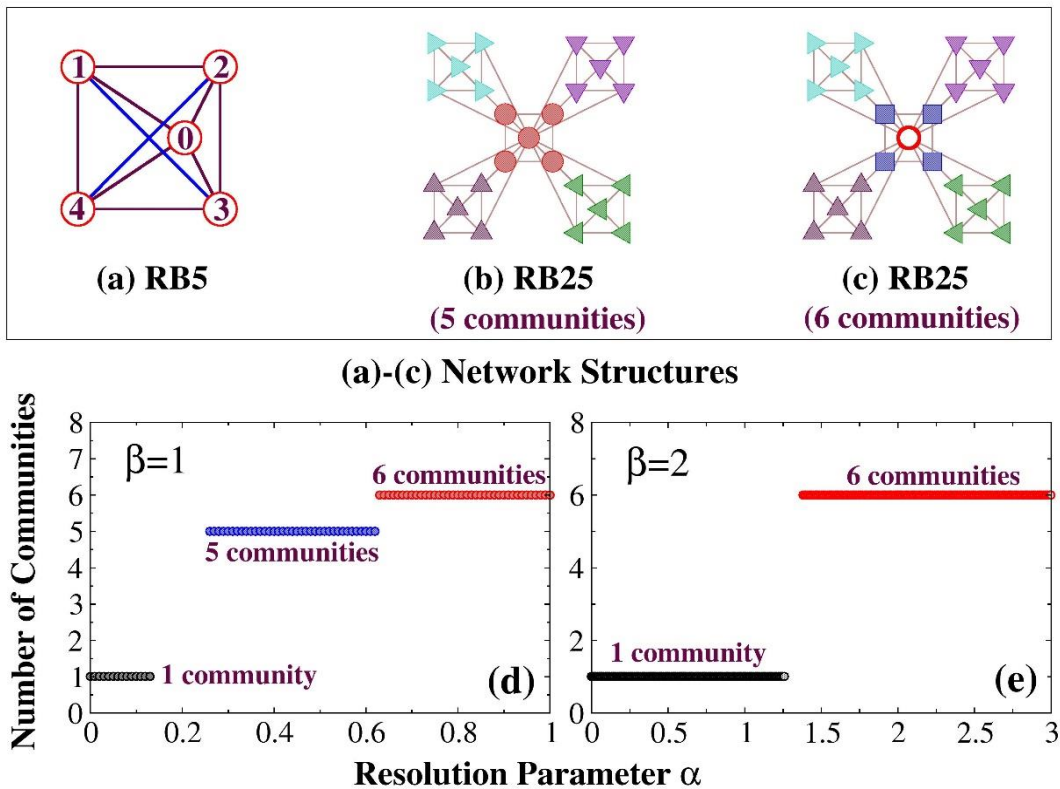

Figure 1. (color online) Communities in the RB25 network. (a) An RB5 network is a complete graph consisting of 5 nodes and 10 edges. We call node 0 the central node, and all other nodes peripheral nodes of the RB5 network. (b) An RB25 network is composed of five RB5 units, one in the center and four on the periphery. Every peripheral node of the peripheral unit is connected to the central node of the central unit, but different peripheral units are not connected to one another. Ideally an RB25 network is expected to be divided into 5 communities: each RB5 unit makes a community. (c) A plausible revision to (b), which divides the RB25 network into 6 communities. Four peripheral RB5 units make four communities, and the central RB5 unit is divided into two communities: the central node makes one community and all peripheral nodes make the other. (d) and (e): Plateaus identified by our method with $\beta=1$ and $\beta=2$. The resolution parameter $\alpha$ varies from 0 to $2\beta-1$, with a stepwise increment $\Delta\alpha = 0.01$. At each resolution, we implement 1000 realizations of the Louvain algorithm, and identify the best-and-unique solutions and plateaus by the strategy described in section 2.3.

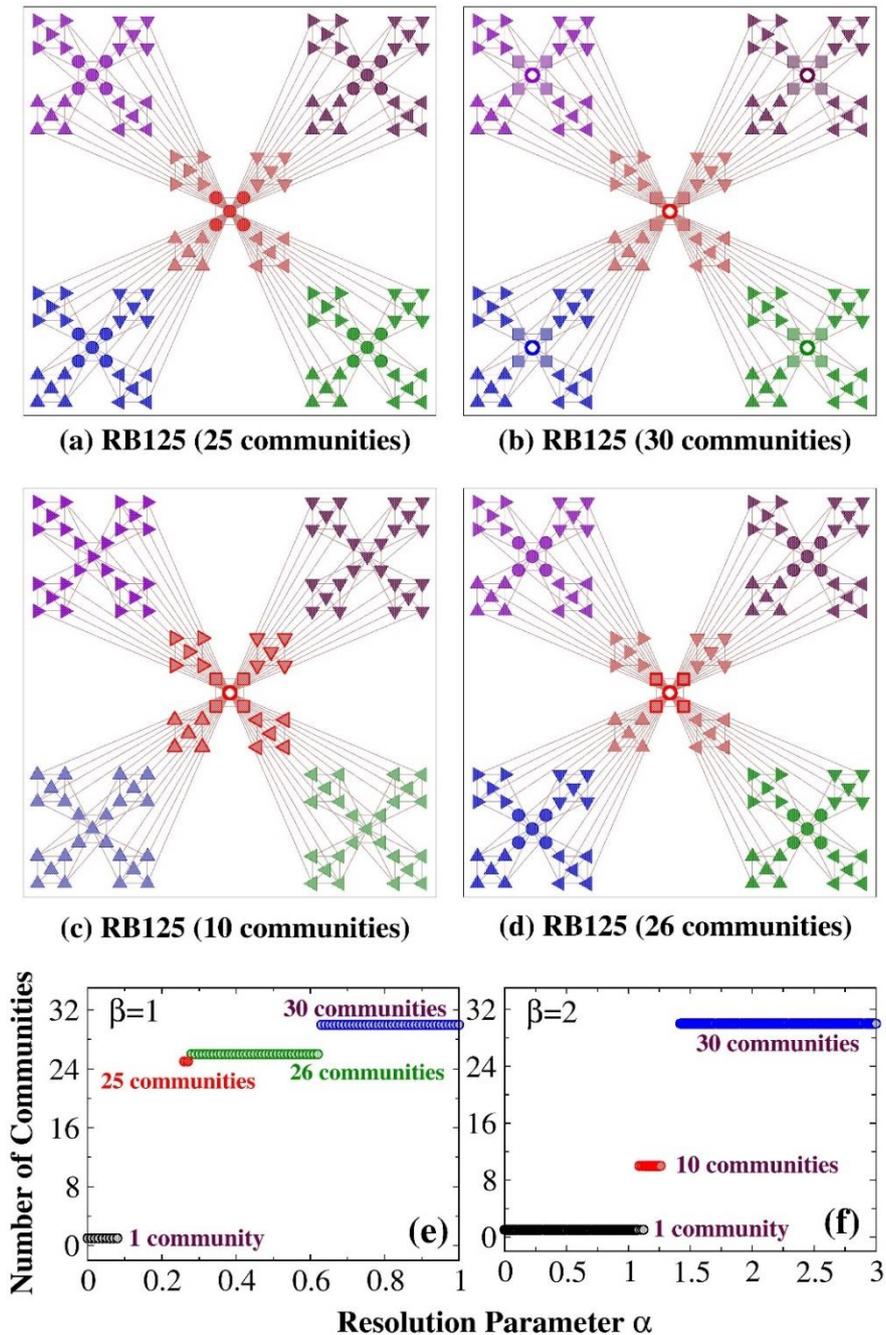

Figure 2. (color online) Communities in the RB125 network. (a)–(d): Four different community structures detected by our method within the RB125 network. In each subfigure, we exhibit different communities with different colors and shapes of nodes (note triangles in different directions also represent different communities). (a) shows a "natural" division for the network: each RB5 unit makes a community; (b) shows our proposed division in table 1 on the lowest community level:

each RB25 unit is divided into 6 communities as in figure 1 (c), and the whole network is divided into 30 communities; (c) shows our proposed division on the second community level, which divides the network into 10 communities; (d) shows an alternative division of the network into 26 communities with a "relaxed stringency" on the lowest community level: only the RB5 unit in the center of the whole network is divided into 2 communities, while all other RB5 units are kept intact. (e) and (f): Plateaus obtained by our method with $\beta=1$ and $\beta=2$. In (e), with $\beta=1$ only the first community level can be detected, but there emerge three different plateaus representing three different divisions for it: 30 communities (our proposed one as in (b)), 25 communities (the "natural" division as in (a)) and 26 communities (the "variant" as in (d)). In (f), with $\beta=2$ all community levels are successfully detected: the first level exhibits only the 30-community division, and the second level only the 10-community division (as in (b) and (c) respectively); we believe these divisions are most stable and "robust" among all potential divisions on the corresponding levels of the RB networks.

| RB5$^n$ Networks | Community Levels ($m$) | | | | | |
|---|---|---|---|---|---|---|
| | 1 (Lowest) | 2 | 3 | 4 | 5 | 6 |
| RB25     ($n=2$) | 6 | 1 | / | / | / | / |
| RB125    ($n=3$) | 30 | 10 | 1 | / | / | / |
| RB625    ($n=4$) | 150 | 50 | 14 | 1 | / | / |
| RB3125   ($n=5$) | 750 | 250 | 70 | 18 | 1 | / |
| RB15625  ($n=6$) | 3750 | 1250 (undetectable) | 350 | 90 | 22 (undetectable) | 1 |

Table 1. Numbers of communities on different community levels in our proposed divisions for the RB networks. Here $m=1$ stands for the lowest community level, which divides the network into smallest communities that cannot split further; $m$ is restricted to be no larger than $n$. Numbers in this table all follow our proposed formula: $(4m+2) \times 5^{n-m-1}$. Our method detects all levels of communities when the network is not extremely large ($n \leq 5$). When $n=6$ (an RB15625 network), the second and fifth levels, which are expected to contain 1250 and 22 communities, turn out to be undetectable by our method.

We first try with $\beta=1$, i.e., the original form of the community fitness function (formula 1). For all RB networks, with $\beta=1$ we only detect two community levels: the highest (but trivial) level that merges the whole network into one single community, and the lowest level that divides the network into smallest communities that cannot split further; *intermediate levels, if exist, are all missed*. On the other hand, with $\beta=1$ we detect different divisions for the lowest level. Except our proposed divisions as listed in the first column of table 1, we also detect the natural divisions, but only for *small* RB networks as RB25 and RB125 (see figure 1 (b) and figure 2 (a)); for larger RB networks, the natural divisions can no longer stand since they violate the customary definitions of community too seriously. Besides, with $\beta=1$ an RB125 network can be divided into 26 communities—this can be done by breaking up only the central RB5 community of each RB125 unit (rather than each RB25 unit) into 2 communities, but preserving all other RB5 communities (as in figure 2 (d)). Compared to the divisions listed in table 1, this alternative division can be understood as a result of a "relaxed stringency," with which an RB5$^n$ network ($n \geq 3$) can be divided into $26 \times 5^{n-3}$ communities; this explains the plateaus representing 26, 130, 650 and 3250 communities that emerge in figure 2 (e) and supplementary figures 1 (b), 2 (a) and 2 (c). Since all these plateaus have similar community sizes ($\leq 5$), we classify them all to the lowest community level. Moreover, with an even furtherly relaxed stringency, where only the central RB5 community of each RB625 unit is broken up into 2 communities, an RB625 network can be divided into 126 communities. Yet such a stringency has been over-relaxed that it only produces a very tiny plateau for the RB625 network in supplementary figure 1 (b), but hasn't been observed anywhere else.

The above results detected with $\beta=1$ are still not satisfying: one major problem is, the highest and lowest community levels have occupied almost all the resolution scales; intermediate levels are seriously compressed and cannot be observed at all. To fix this problem, we take $\beta \geqslant 2$. It turns out that with $\beta \geqslant 2$ our method discovers *all community levels* including the intermediate ones for the RB networks; see figures 1 (e), 2 (f) and supplementary figures 1 (c) and 2 (b) for the plateaus obtained with $\beta=2$. With $\beta>2$ we simply get similar results. As observed, for RB networks with no more than 5 levels, our method *accurately and exclusively* recovers the divisions suggested in table 1. As for the potential variants due to relaxed stringencies, in some realizations we did detect some of them as "local best solutions." Yet globally (i.e., over all realizations), these variants are not best-and-unique, thus are discarded by our filtering strategy. In contrast, the divisions suggested in table 1 perform more robustly, and our filtering strategy accurately hits on these "robust divisions." On the other hand, our results also confirm that the scaling factor $\beta$ does rescale the community fitness function effectively, which facilitates our detections on multilevel community structures.

Yet our method also has a limit. We notice that with the increase of $\beta$, the resolution ranges for different community levels span differently: the highest and lowest levels always occupy a majority of the resolution scales, while intermediate levels only emerge within a limited region (when $\beta=2$, it displays as $1<\alpha<1.6$). For RB networks deeper than 5 community levels, some intermediate levels will be compressed and cannot be detected by our method. For example, for an RB15625 network, in supplementary figure 2 (d) plateaus for the second and fifth community levels, which are expected to represent 1250 and 22 communities, are both missing. Within the resolution scales where these community levels are expected to emerge, the Louvain algorithm fails to converge to a best-and-unique solution. Furtherly increasing $\beta$ does not solve this problem. As shown in supplementary figure 3, with the increase of $\beta$, resolution scales for intermediate community levels do not expand remarkably. Within the scope of this paper, the limit of our method is to detect up to five community levels for the RB networks; to detect more community levels, an improved method that rescales the resolution ranges for different community levels more evenly, should be worth studying in the future.

Lastly, we show the performances of some earlier methods on the RB networks for comparison. The RB networks have two distinctive features: hierarchical and symmetrical. Methods without a tunable resolution parameter are not expected to detect multiple levels of communities from an RB network, but they can (or should) be expected to detect at least communities of one level perfectly. In supplementary figure 4, we show the communities detected for an RB125 network by two well commended methods: (1) the standard modularity $Q$ proposed by Newman [39], optimized through a Louvain algorithm, and (2) Infomap [21]. We find both these methods inevitably produce divisions with *randomness* through breaking the symmetry of the network: all communities in supplementary figure 4 are detected in a random manner. For example, in (a), the central RB25 unit is divided into three communities by *randomly* combing two of its four peripheral RB5 units with the central unit, and each of the peripheral RB25 units is divided into two communities by *randomly* choosing one of its peripheral RB5 units as an individual community—such a division could be the best in terms of modularity, but it is definitely not unique. In the context that network nodes are distinguishable, considering the symmetry of the network, we can easily calculate there exist 1536 divisions that are equivalent to supplementary figure 4 (a); within 1000 independent realizations, almost every single realization suggests a different division. Similar is the division given by Infomap in (b), which has

256 equivalents. In contrast, community levels detected by our method, as in table 1, retain both the hierarchy and symmetry of the network. Therefore, communities detected by our method should be more reasonable and significant than those detected by modularity $Q$ and Infomap.

As for methods with tunable resolutions, we choose the multiresolution version of modularity $Q$ generalized by Reichardt and Bornholdt [24]; other methods as those in [3, 26] essentially perform identically [33]. The multiresolution modularity tunes the resolution through a parameter $\gamma$:

$$Q_\gamma = \sum_{\mathcal{G}} \left[ \frac{k_{in}^{\mathcal{G}}}{2m} - \gamma \left( \frac{k^{\mathcal{G}}}{2m} \right)^2 \right] \; ; \qquad \text{(Formula 4)}$$

Here $m$ represents the total number of edges within the whole network; $k_{in}^{\mathcal{G}}$ and $k^{\mathcal{G}}$ represent the in-degree and total degree of community $\mathcal{G}$, and the modularity is summed over all communities. In supplementary figure 5, we exhibit the plateaus detected by $Q_\gamma$ for three RB networks. Only for the smallest RB125 network, $Q_\gamma$ detects all three community levels. For larger networks such as RB625 and RB3125, $Q_\gamma$ detects no more than two levels of communities for each of them. Therefore, our method outperforms previous multiresolution methods on deep hierarchical networks—as we will suggest in the Discussion, the scalability of our community fitness function plays an important role.

### 3.2 On the heterogeneous Lancichinetti-Fortunato-Radicchi (LFR) benchmark networks

Benchmark networks with implanted communities generated by early methods, including the traditional stochastic block model (SBM) [15, 16], and the GN benchmark [2], differ substantially from their real-world counterparts: real-world networks typically have heterogeneous distributions of node degree and community size [43]. The LFR benchmark network was proposed to address the issue. Its node degrees and community sizes follow different power-law distributions, and may both span more than one order of magnitude. Each node is planted into a community: it shares a fraction of 1-$\mu$ of its connections with the other nodes of the same community, and the rest fraction $\mu$ with nodes in other communities; $\mu$ is called a *mixing parameter* [44]. Large values of $\mu$ weaken the validity of the implanted communities. Especially, when $\mu \geqslant 0.5$, the implanted communities violate the customary definitions of community in both a strong sense and a weak sense, which respectively require $k_{in}>k_{out}$ for every node, or $\Sigma k_{in}>\Sigma k_{out}$ for every community [41]. However, many community detection methods keep recovering the implanted communities perfectly even when $\mu$ is near 0.8; the reasonability is based on the fact that most nodes in the network still retain more connections within its own community than sharing connections with *any other* communities, as suggested by [45]. In previous literature, it is general to employ a normalized mutual information (NMI) [46] to evaluate the consistency between the detected communities and the implanted ones; NMI equals to 1 indicates that they agree with each other perfectly.

In [33], the authors argued on an LFR network, when the community sizes vary enormously, all previous multiresolution methods lose their effectiveness. Multiresolution methods are reported to outperform other methods only because the community sizes used in their tests were too close to one another. In larger networks with more heterogeneous community sizes, multiresolution methods all fail to detect the expected communities even when $\mu$ is far below 0.5. In contrast, Infomap, being lack of a tunable resolution parameter, performs much better (see figures 6-9 in [33]).

We test the effectiveness of our method under the same conditions: on LFR networks built with exactly the same parameters as those in figures 6-9 of [33], we exhibit in figures 3 & 4 the NMIs between the implanted communities and the communities detected by our strongest (shown in red squares) and second strongest (shown in blue circles) plateaus with different values of $\mu$. Networks in figure 3 are relatively small and contain communities of similar sizes, while networks in figure 4 are larger and more heterogeneous. Since the LFR networks contain no multilevel structure, for each network we only run 1000 realizations of the Louvain algorithm with $\beta=1$; $\beta \geqslant 2$ simply yields the same results. For comparison we show in the same figures the NMIs for communities detected by Infomap (shown in black triangles). We use $\mu_1$ in each figure to indicate the threshold at which the NMIs of our strongest plateaus start to deviate from perfection: when $\mu \leqslant \mu_1$, the NMIs always equal to 1. When $\mu > \mu_1$, our strongest plateaus detect the whole network as one community, thus the NMIs suddenly decrease to 0. But our second strongest plateaus still recover the implanted communities perfectly and retain high levels of NMIs, till $\mu$ becomes really large. Similarly, we use $\mu_2$ to indicate the same threshold for Infomap: when $\mu > \mu_2$, the whole network is detected as one single community. Since for Infomap, there is no "secondary solutions," thus when $\mu > \mu_2$, the NMI always equals to 0.

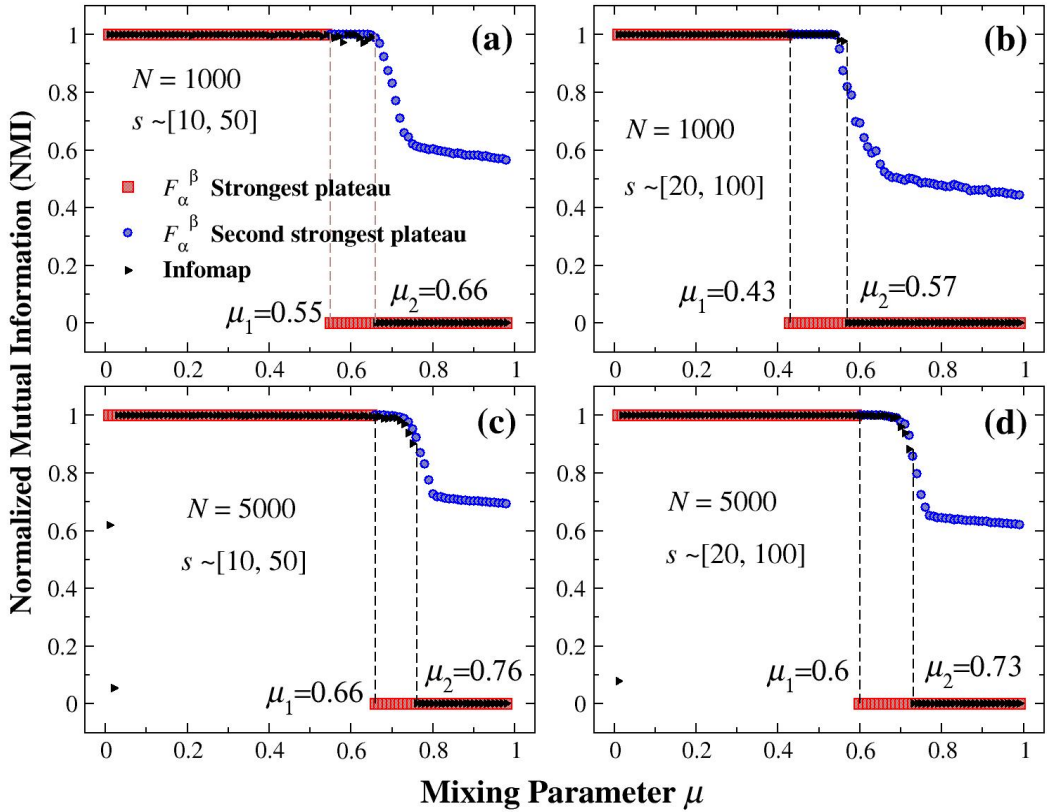

Figure 3. (color online) Normalized mutual information (NMI) between the communities detected in relatively small LFR networks by our method and Infomap against the implanted communities with varying $\mu$. The following network parameters are shared among all subfigures: average node degree $<k> = 20$, maximum node degree $k_{max}=50$, power-law distribution of the node degree $k$: $f(k) \sim k^{-2}$, and of the community size $s$: $g(s) \sim s^{-1}$. Other parameters including the network size $N$ and the range of community size $s$ are labelled in each subfigure.

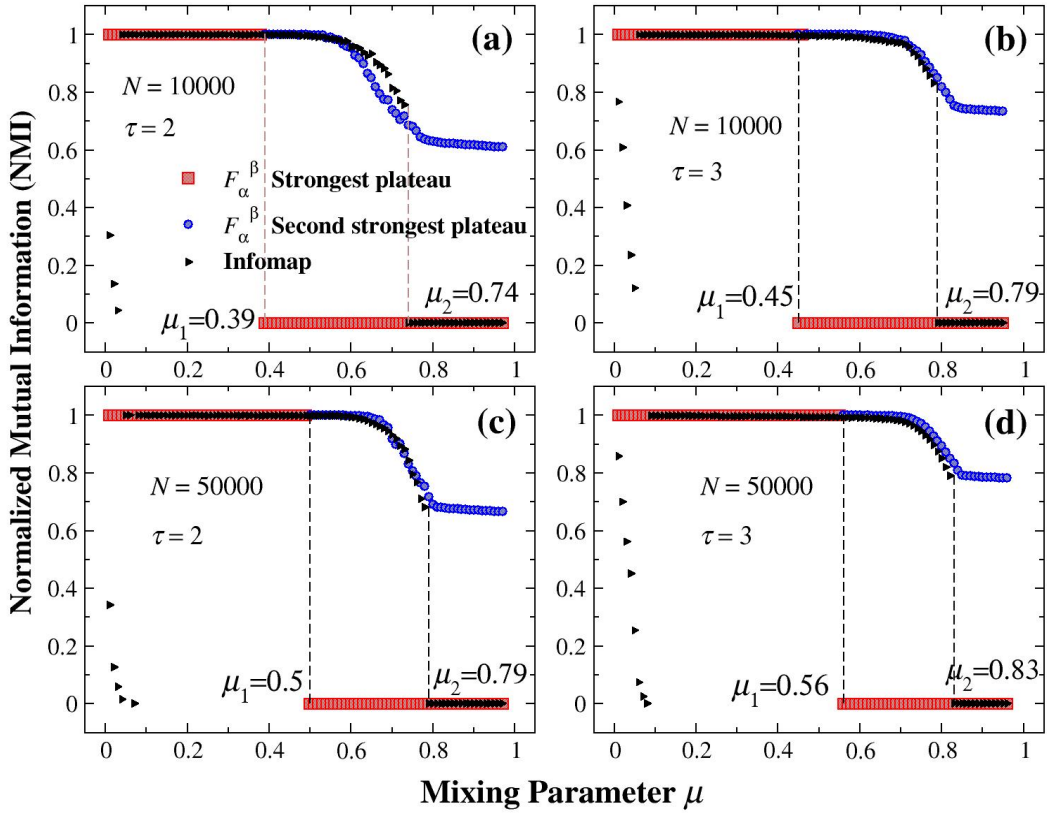

Figure 4. (color online) Normalized mutual information (NMI) between the communities detected in larger and more heterogeneous LFR networks by our method and Infomap against the implanted communities with varying $\mu$. The following network parameters are shared among all subfigures: average degree $<k>=20$, maximum degree $k_{max}=100$, power-law distribution of the node degree $k$: $f(k) \sim k^{-2}$, range of community sizes $s \sim [10, 100]$. Other parameters including the network size $N$ and the power-law exponent of the community size distribution $\tau$ are labelled in each subfigure.

Previous multiresolution methods tested in [33] mostly perform well on the "small" networks in figure 3—but they all perform much worse on the large and heterogeneous networks in figure 4. In contrast, our method, being also a multiresolution method, recreates the implanted communities for all networks in figures 3 and 4 as perfectly as Infomap within the range $\mu \leqslant \mu_2$; their NMI curves roughly overlap with each other. More specifically, when $\mu \leqslant \mu_1$ or $\mu > \mu_2$, our strongest plateaus yield exactly the same results as Infomap; when $\mu_1 < \mu < \mu_2$, our second strongest plateaus and Infomap recover the implanted communities to the same extent of perfection. Compared to previous methods of multiresolution that are tested in [33], our method performs more robustly on large heterogeneous networks. It seems to have overcome the resolution limit problem caused by network heterogeneity. As we will suggest in the Discussion, such an outperformance too, can be attributed to the scalability of our community fitness function.

Besides, our best-fit NMIs shown in figures 3 & 4 are *not* artificially selected: we clearly know our strongest plateaus suggest the most significant results—in case the strongest plateaus suggest to "trivially" merge the whole network into one community, the second strongest plateaus may remain suggestive. This means that our best-fit solutions are detected *automatically*, without the need of knowing any information about the implanted communities. In contrast, previous methods including that in [26], rely on the *a priori* knowledge to judge from a mess of detection results which ones are the best—without knowing the implanted communities, there is no clue to pick out the results with the "strongest correlations" [26]. Our automatic community detection should give the credit to the filtering strategy suggested in section 2.3, which guarantees the stability, as well as significance, of all our final outputs.

**3.3 Applications to real-world networks**

Unlike synthetic networks, real-world networks have no implanted communities. Instead, there are observed discrete-valued node attributes, or metadata, being customarily used as a proxy of the ground truth [43, 47]. In early literature, it was common to validate the effectiveness of a community detection algorithm by its success in recovering the metadata: if the detected communities correlate with the metadata, then one can reasonably conclude that the corresponding algorithm is promising to work effectively in practice—but its opposite has been realized more recently as being not true, i.e., failing to fit the metadata does not necessarily signify the failure of the algorithm [47]. Since synthetic (i.e., artificial) networks may not be representative of naturally occurring interactions, applications to real-world networks are still worth checking for community detection algorithms.

Considering the recent viewpoint on metadata [47], we do *not* intend to validate the reliability of our method by its performance on real-world networks, or not merely that. We do not manage to fit our results unconditionally to the metadata, instead we try to put some insight into the differences between them. The scientific value of a method is as much defined by the way it fails as by its ability to succeed [47]; a reasonable but different outcome to the metadata can hopefully reveal a different aspect of the network structure. In this section, we investigate three real-world networks: Zachary's karate club network [48], Lusseau *et al.*'s dolphins social network [49], and the American college football network [2]. We detect communities for these networks with our method, and compare our results with the metadata or "standard divisions" given by previous literature.

Figure 5 shows the community structures within the karate club network. This network consists of 34 nodes representing 34 members of a karate club; connections between nodes imply consistent interactions between the corresponding members outside the club. Due to a disagreement between the club president (John A.) and a part-time instructor (Mr. Hi), the original club later split into two parts, the officers' club and Mr. Hi's; members of the original club also diverge to follow their own favorite leader (see the communities split by the dashed line in figure 5 (a), which are referred to by the "metadata division" in the next paragraphs). Community detection methods in previous literature attempted to recreate such a division by various models; among them the result given by Newman and Girvan [7] is often taken as a "standard" division for the karate club network.

Applying our method to the karate club network, we find a distinctive difference between real-world networks and synthetic networks—that is, in terms of the number of communities, synthetic networks are implanted with discrete levels of communities, while real-world networks may display more "continuous" community levels. As shown in figure 5 (c) and (d), with varying resolution, our method detects 11 levels of communities with $\beta=1$, and 10 levels with $\beta=2$ in the karate club network; the numbers of communities include every value from 1 to 12. Surprisingly, these community levels are all stable and unique, whose topologies are exhibited in supplementary figure 6, forming roughly a hierarchical structure, with only minor "reassembling" of communities in the 5- and 6-community levels (detected with $\beta=1$ only, denoted by the dashed rectangles in supplementary figure 6). Among all these levels, our 2-community division differs from the metadata—but it is fully consistent with the division suggested in [50, 51], which has also been known as the only way to divide the karate club network into communities all defined in a strong sense [52]. Except the 2-community level, communities of all the rest levels can be properly combined to recreate the metadata division. For example, in figure 5 (a) and (b), we exhibit our 4- and 5-community divisions respectively by nodes of different shapes. Obviously, combining circles and octagons in both (a) and (b) roughly recovers Mr. Hi's club, while combing squares and hexagons (and also diamonds in (b)) roughly recreates the officers'.

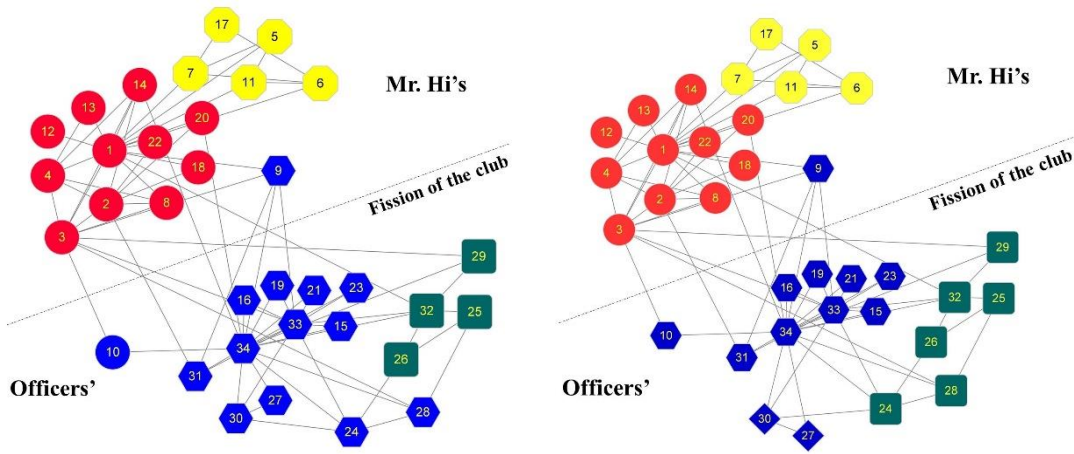

(a) Divisions for the karate club network given by our 4-community level and Newman-Girvan [7].

(b) Divisions for the karate club network given by our 5-community level and Medus *et al.*'s simulated annealing approach [53].

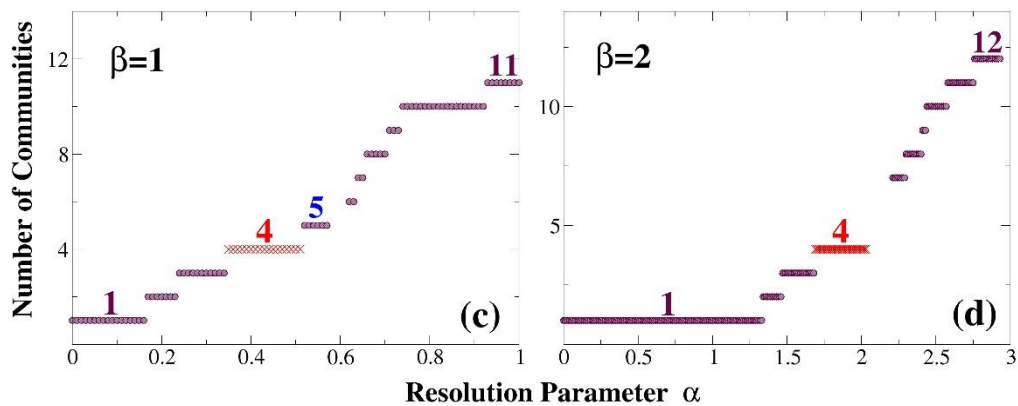

Figure 5. (color online) Communities detected in the karate club network. (a) Divisions given by our 4-community level and Newman-Girvan [7]. Our division is shown by nodes of different shapes (octagon, circle, square, and hexagon), while Newman and Girvan's is by nodes of different colors (yellow, red, blue and green). (b) Divisions given by our 5-community level and Medus *et al.* [53]. Our division is shown by nodes of different shapes (octagon, circle, square, hexagon and diamond), while Medus *et al.*'s is by nodes of different colors (yellow, red, blue and green). The dashed line drawn in both (a) and (b) divides the network into two parts, corresponding to a fission which had actually happened to the club. (c)-(d) Plateaus detected by our method from 1000 realizations of the Louvain algorithm at each resolution with $\beta=1$ and $\beta=2$. Numbers above the plateaus indicate the corresponding numbers of communities. Figure 6 (c)-(d) and figure 7 (b)-(c) in the following part of this paper are also drawn in the same way.

Now we compare our divisions with previous ones. As same as previous methods [7, 53], our method also subdivides the metadata division into more communities. As shown in figure 5 (a), our 4-community division is largely consistent with the division given by Newman and Girvan through their shortest path betweenness method [7], except one node: node 10. We notice that node 10 has only two neighbors: node 3 joined Mr. Hi's club, and node 34 joined the officers'. It seems difficult to determine which choice for node 10 should be better than the other based on the network structure. On the other hand, compared to the metadata division, both our result and Newman-Girvan's have misclassified node 9 to the officer's club, since node 9 evidently has more connections to the officers' club than to Mr. Hi's. Actually, in the original literature of Zachary's [48], there are two metadata attributes recorded: the political leaning of each of the members and the faction they finally joined after the club fission. Previous literature on community detection only used the latter to evaluate the results, so that node 9 is almost always mislabeled. Considering the metadata on the political leaning of members (see table 1 in [48]), node 9 was actually a weak supporter of the officer, but he chose to join Mr. Hi's club only for the convenience of a coming exam for his black belt. While node 10 was identified as a member of no faction, who may have probably chosen the faction randomly. As suggested by [47], the detected communities and the metadata may capture different aspects of the network structure, thus some misclassification of nodes may also provide worthy information about the network.

Alternatively, the simulated annealing approach applied by Medus *et al.* results in a different division for the karate club network [53]. In figure 5 (b), we demonstrate that Medus *et al.*'s division is fully consistent with our 5-community division, if we combine the smallest community (diamonds, i.e., nodes 27 and 30) with a larger community (squares). And on this community level, node 10 is correctly classified (to the officer's club). Our result suggests that Newman-Girvan's division and Medus *et al.*'s, looked different, are probably results observed at different resolutions: they represent different community levels, and are basically both correct.

Next we move to the dolphins' social network (hereafter we call it the "dolphins' network" for short). The dolphins' network was compiled by Lusseau and his collaborators from seven years of filed studies on a bottlenose dolphins' society living in Doubtful Sound, New Zealand [49, 54]. To our knowledge, the first version of this network was established in [54], including 40 individuals of dolphins. After that, an extended version including 62 nodes and 159 edges was published in [49], which is the "dolphins' network" widely studied by community detection literature including this paper. Nodes of the network represent the population of dolphins, while edges reflect associations between dolphin pairs occurring more often than expected by chance [49, 55]. Newman and Girvan firstly divided this network into two communities in [7], which allegedly correspond to a "known division" of the dolphins' society. However, as far as we know, such a "known division" was not included in the metadata recorded in [49, 54], thus previous literature actually took Newman and Girvan's division as a standard division. The larger community of Newman-Girvan's can be further divided into four smaller communities [7], as visualized in figure 6 (a). Such divisions have been cited later by both Newman and Lusseau [55, 56]. In [55], the smallest community containing only two nodes (Zipfel and TSN83, i.e., the purple community at the top of figure 6 (a)) was merged into a larger community (the red community right in the middle of Group 1), so that the total number of communities decreased to four [55, 56].

Applying our method to the dolphins' network, we get similar results as the karate club network. Among the multiple community levels, our 2-community level exhibits a strongest plateau (except the trivial single-community level, see figure 6 (c) and (d)). The corresponding network division is visualized in figure 6 (b) by nodes of different shapes (rectangles and ellipses). Compared to the standard division given by Newman and Girvan, our 2-community division misclassifies only one node, SN89, among all 62 nodes, to a different group. We notice that SN89 has only two connections: one to SN100, the central node with the highest betweenness [55] in Group 1, and the other to Web, an individual in Group 2. It was said that the "known division" between the two groups of dolphins was due to a temporary leave of SN100: interactions between the two groups were restricted while SN100 was away and became more common when it reappeared [7, 55]. We argue that when SN100 was away, presumably the interaction between SN89 and Group 1 should also be cut off—however its interaction with Group 2 can be maintained through the connection to Web. Therefore, on the 2-community level, our classification for SN89 should be more reasonable, which fits the ground truth better.

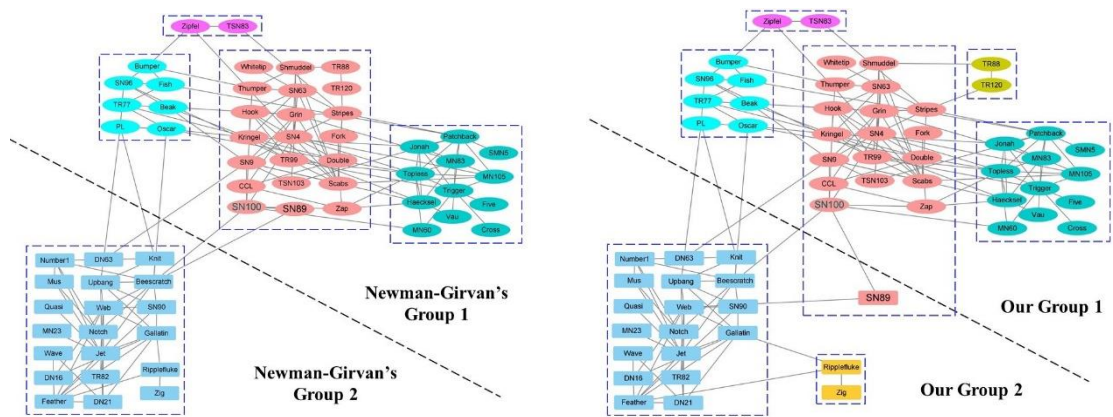

(a) Newman and Girvan's 2-community and 5-community divisions for the dolphins' social network

(b) Our 2-community and 7-community divisions for the dolphins' social network

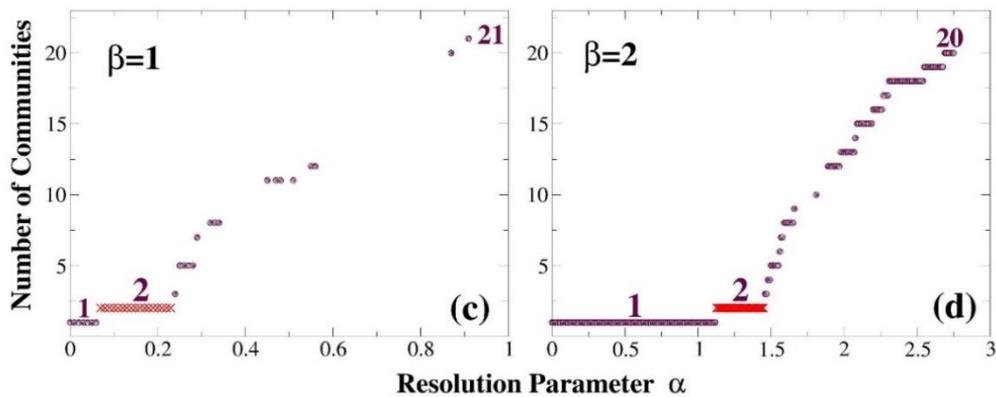

Figure 6. (color online) Communities in the dolphins' network. (a) Newman and Girvan's divisions. The whole population is firstly divided into two major groups (rectangles and ellipses, separated by the dashed line), then ellipses can be further divided into four small communities shown in different colors (cyan, purple, red and turquoise). (b) Our divisions. The whole population can be divided into either two communities (rectangles and ellipses, on opposite sides of the dashed line), or seven small communities shown in different colors (blue, orange, cyan, red, turquoise, green and purple). For better visibility, in both (a) and (b) we enclose the small communities shown in different colors in dashed boxes. (c)-(d) Plateaus drawn in the same way as figure 5 (c)-(d).

But shifting to a higher resolution, the result is different. In figure 6 (b), we exhibit our 7-community division by nodes of different colors; the corresponding communities are also enclosed in different boxes for better visibility. It turns out that two of our smallest communities (yellow and green, on the bottom middle and upper right of figure 6 (b)) can be merged into the blue and red communities to perfectly recreate Newman and Girvan's 5-community division. And on this level, node SN89 has also been reclassified to the "right" group.

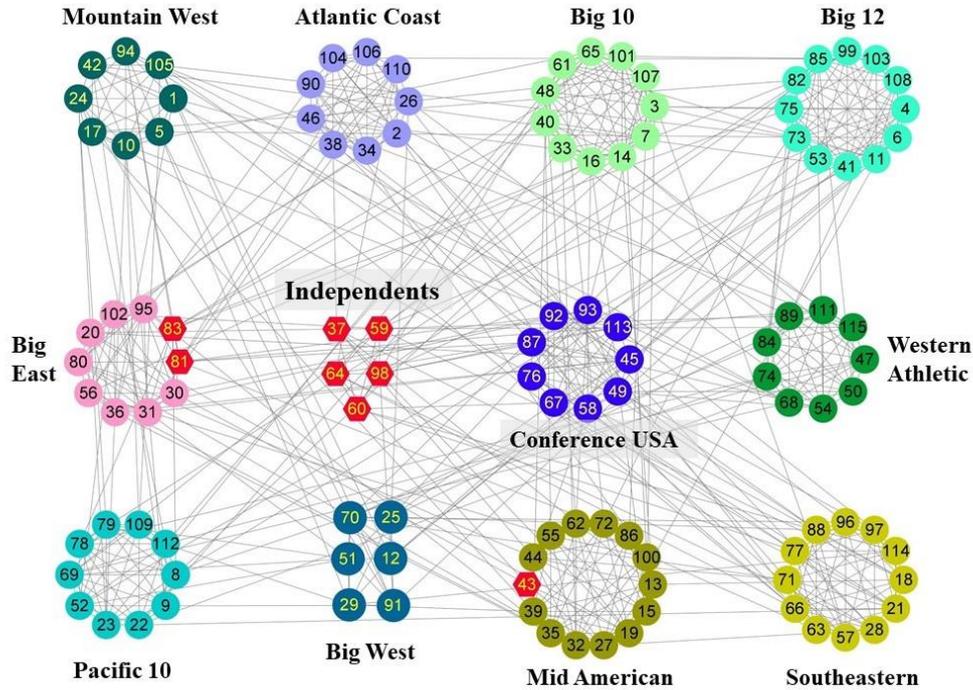

(a) **Communities in the American college football network.**

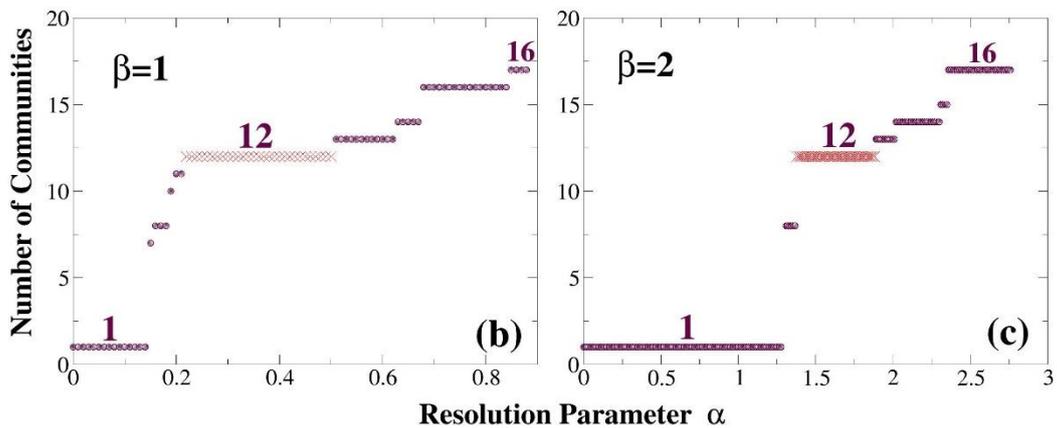

Figure 7. (color online) Communities in the American college football network. (a) Structure of the network: 115 nodes represent 115 American college football teams, and edges between them denote scheduled games between these teams. Except 8 independent teams (denoted by the red hexagons), other teams are all affiliated with 11 different conferences. We show teams of different conferences by nodes of different colors, and display our 12-community division by naturally separated clusters. It turns out that our division perfectly recovers the members of all 11 conferences, and 5 independent

teams out of 8 are recognized by an individual community (annotated as "Independent"). (b)-(c) Plateaus drawn in the same way as figure 5 (c)-(d).

The last network we study is the American college football network (next for short we call it the "football network"); it was constructed from the schedule of the Division I games of the 2000 season of United States college football [2]. This network consists of 115 nodes representing 115 college football teams, distinguished by their college names. Among all teams, 107 were affiliated with 11 different conferences each containing 6 to 13 teams, and the rest 8 teams were independent of any conference. Edges of the network represent scheduled games between the connected teams during the 2000 season, which turned out to be much more frequent between teams of the same conference than between those of different conferences. Since Girvan and Newman firstly recreated the conference assignments correctly for most teams with their algorithm in 2002 [2], the football network has been cited and investigated repeatedly in community detection literature. However, in 2010 Evans pointed out there was a serious error in Girvan and Newman's metadata recorded in figure 5 of [2]: the conference assignments for those teams seemed to be collected during the 2001 season rather than the 2000 season [47, 57]. In figure 7 (a), we exhibit the conference assignments corrected by Evans in Appendix C2 of [57] with nodes of different colors; specially, independent teams are denoted by red hexagons. And we also annotate the corresponding name of the conference beside each group of nodes. For comparison, we exhibit in supplementary figure 7 the metadata of Girvan and Newman's [2]; validity of the metadata has been demonstrated in [57].

With our method, we also detect multilevel communities in the football network, see figure 7 (b) and (c) for the plateaus. Among them, the strongest plateau suggests a 12-community division, which is naturally displayed in figure 7 (a) with edges inside communities being shorter than those between different communities. Obviously, our division perfectly recovers the members for all 11 conferences: teams of the same conference are all classified into the same community. As for the 8 independent teams, 5 of them have been put into an individual community ("Independents" in figure 7 (a)), and the rest 3 teams are assigned to two conferences that they played most their games with. Girvan and Newman's division agrees with ours, except only one node (node 37, representing team "Central Florida") is assigned to conference "Mid America" [2], which to our viewpoint, is only a minor difference.

Apparently, detection "errors" observed in supplementary figure 7, as well as those reported by Girvan and Newman in [2], are both due to the errors in the metadata [57], rather than the failure of the community detection algorithms [47].

## 4. Discussion

Above we have validated the effectiveness of our method on synthetic benchmark networks, including the hierarchical RB networks and the heterogeneous LFR networks. We also investigated its applications on real-world networks, and exhibited the consistency between our results and the metadata. The outperformance of our method can be attributed to two of its distinctive features: (1) the scalability of the community fitness function, and (2) the stability of the outputs. Next we make some discussions on the features of our method.

**Scalability of the community fitness function**

The scalability of our community fitness function (formula 2) originates from two aspects. First, its original form (formula 1) as introduced in [20] is by design more scalable than other community quality functions, for example, the standard modularity $Q$ proposed by Newman [17], which is generally reformulated as [3, 13, 28, 33]

$$Q = \sum_{\mathcal{G}} Q^{\mathcal{G}} = \sum_{\mathcal{G}} \left[ \frac{k_{in}^{\mathcal{G}}}{2m} - \left( \frac{k^{\mathcal{G}}}{2m} \right)^2 \right].$$

Here $m$ is the total number of edges within the whole network; $Q^{\mathcal{G}}$, $k_{in}^{\mathcal{G}}$ and $k^{\mathcal{G}} = k_{in}^{\mathcal{G}} + k_{out}^{\mathcal{G}}$ are the modularity, in-degree and total degree of community $\mathcal{G}$. Obviously, the value of $Q^{\mathcal{G}}$ depends heavily on the community size: in an LFR network, since $k_{in}^{\mathcal{G}} / k^{\mathcal{G}} = 1 - \mu$ ($\mu$ is the mixing parameter),

$$Q^{\mathcal{G}} = \frac{k_{in}^{\mathcal{G}}}{2m} - \left( \frac{k^{\mathcal{G}}}{2m} \right)^2 = \frac{k_{in}^{\mathcal{G}}}{2m} \left[ 1 - \frac{k^{\mathcal{G}}}{2m(1-\mu)} \right].$$

In large networks, it can be expected that $2m \gg k^{\mathcal{G}}$, thus $\frac{k^{\mathcal{G}}}{2m(1-\mu)} \approx 0$, so that $Q^{\mathcal{G}} \approx \frac{k_{in}^{\mathcal{G}}}{2m} \propto k_{in}^{\mathcal{G}}$.

This reflects a fact that the modularity $Q^{\mathcal{G}}$ increases almost linearly with the *community size* (here without loss of generality, we measure the community size by the in-degree rather than the number of nodes). Large gaps of $Q^{\mathcal{G}}$ exist between communities of different sizes in a heterogeneous network, which inhibits simultaneous detections on communities of distinct sizes. In contrast, the community fitness function (formula 1)

$$f^{\mathcal{G}} = \frac{k_{in}^{\mathcal{G}}}{(k^{\mathcal{G}})^{\alpha}} = (1-\mu)^{\alpha} (k_{in}^{\mathcal{G}})^{1-\alpha} \propto (k_{in}^{\mathcal{G}})^{1-\alpha}.$$

Since $\alpha > 0$, apparently $f^{\mathcal{G}}$ increases not as fast as $Q^{\mathcal{G}}$ with the community size: it tends to narrow the gap between large and small communities. As a result, in a heterogeneous network, communities having close densities of inner connections but far different sizes can be simultaneously identified by the community fitness function $f^{\mathcal{G}}$, but not by the modularity function $Q$. The multiresolution version of $Q$ proposed by Reichardt and Bornholdt [24] does not solve the problem: it introduces a resolution parameter $\gamma$ as in formula 4:

$$Q_{\gamma} = \sum_{\mathcal{G}} \left[ \frac{k_{in}^{\mathcal{G}}}{2m} - \gamma \left( \frac{k^{\mathcal{G}}}{2m} \right)^2 \right].$$

However, since $\left( \frac{k^{\mathcal{G}}}{2m} \right)^2$ is a minor item of the formula, $\gamma$ is not effective enough to rescale $Q_{\gamma}$ and overcome the resolution limit problem [33]. Similar problem also holds for a majority of previous multiresolution methods, including the popular Hamiltonian-based Potts models [23, 24, 25, 26, 27]. That explains why previous multiresolution methods perform poorly on heterogeneous networks, as argued by [33], while our method in this paper have outperformed all of them (see section 3.2).

The second origination of the scalability is the scaling factor $\beta$: we can say $\beta$ makes the fitness function "rescalable." The original community fitness function has $\beta$ fixed to 1; correspondingly, the varying range of the resolution parameter $\alpha$ is between 0 and 1. According to our estimation in the supplementary material, such a varying range is a "relevant" scale of resolution, within which both the merging of small communities and the splitting of large ones are restricted. Therefore, in multilevel networks, only the lowest community level can be detected, while all intermediate levels are omitted. In contrast, when $\beta>1$, it rescales the whole resolution range by a multiple $2\beta$-1, which amplifies the resolution scales of different community levels, and effectively enables our detections on all levels of communities. As a result, for the RB hierarchical networks, our method successfully detects up to five levels of communities with $\beta=2$, which to our knowledge, has not been done by any other methods reported in previous literature.

Yet the deficiency of our scaling factor $\beta$ is that it rescales the resolutions unevenly: comparing to the lowest and highest (but trivial) levels, expansions for the resolution scales of the intermediate community levels are relatively minor. As a result, for networks having too many community levels, our method fails to detect some of them. Additionally, when the network size is too large, it becomes more and more difficult for the original Louvain algorithm to converge to a best solution. Improved methods and algorithms are to be studied in the future.

**Stability of the outputs**

Stability of the outputs is mainly due to our strict stringencies of both defining and identifying the plateaus. It has been popular to rank the significance of outputs by the persistence of plateaus in previous literature. However, as we have argued in the Introduction, if the term "plateau" was only loosely defined, which cannot guarantee "one plateau one topology," the related ranks are not surely trustworthy. Although we believe most plateaus in previous literature did have consistent topologies, such an important issue hasn't been stated even once.

In this paper, we suggested a strict stringency that requires not only "one plateau one topology," but also a "best-and-unique" solution for each relevant resolution. With this stringency, we removed the above suspicions on plateaus, and also rejected unstable results of detection. Here we raise just a simplest example: for an RB25 network, resolution scales exhibited in figure 8 (e)-(f), i.e., $0.14 \leqslant \alpha \leqslant 0.25$ for $\beta=1$, and $1.27 \leqslant \alpha \leqslant 1.37$ for $\beta=2$, are both unstable. Within these resolution scales, best solutions detected by the Louvain algorithm do not have unique topologies. For instance, figure 8 (a)-(d) show four different 4-community divisions for an RB25 network. In the context that network nodes are distinguishable, due to the symmetry of the network, these different divisions all have the same value of $F_\alpha^\beta$. In certain resolution scales, they can be all detected as best solutions—but not unique. To our viewpoint, divisions in figure 8 (a)-(d) are nondeterministic: the central RB5 unit is combined with a *randomly chosen* peripheral unit. That is to say, these detected communities result from *random convergences* of the Louvain algorithm, and cannot be expected to be informative on any attributes of the network. Supplementary figure 4 also reflects the same issue. In figure 8 (e)-(f), we split the unstable resolution scales to subintervals; each subinterval is tagged with the number of communities in the "best-but-non-unique" solutions detected therein. Similar phenomena also exist in asymmetrical networks; in the LFR and real-world networks, we also observe large numbers of "best-but-non-unique" solutions. Our method is designed to *automatically* remove such solutions. As a result, our diagrams of plateaus are very *clean*; all plateaus are stable, and relevant to known community structures. They are largely interpretable, without the need of any artificial or arbitrary selection. In contrast, previous methods did not require uniqueness of the solutions; plateaus for the same networks detected by these methods are much more *redundant*. Small plateaus emerge in the transitional regions between large plateaus; see figure 1 in [3], figure 2 in [26], figure 3 in [27], figures 2&3 in [37], and so on. These small plateaus are mostly uninterpretable; one has to rely on the *a priori* knowledge to judge, or even arbitrarily select the favorite results.

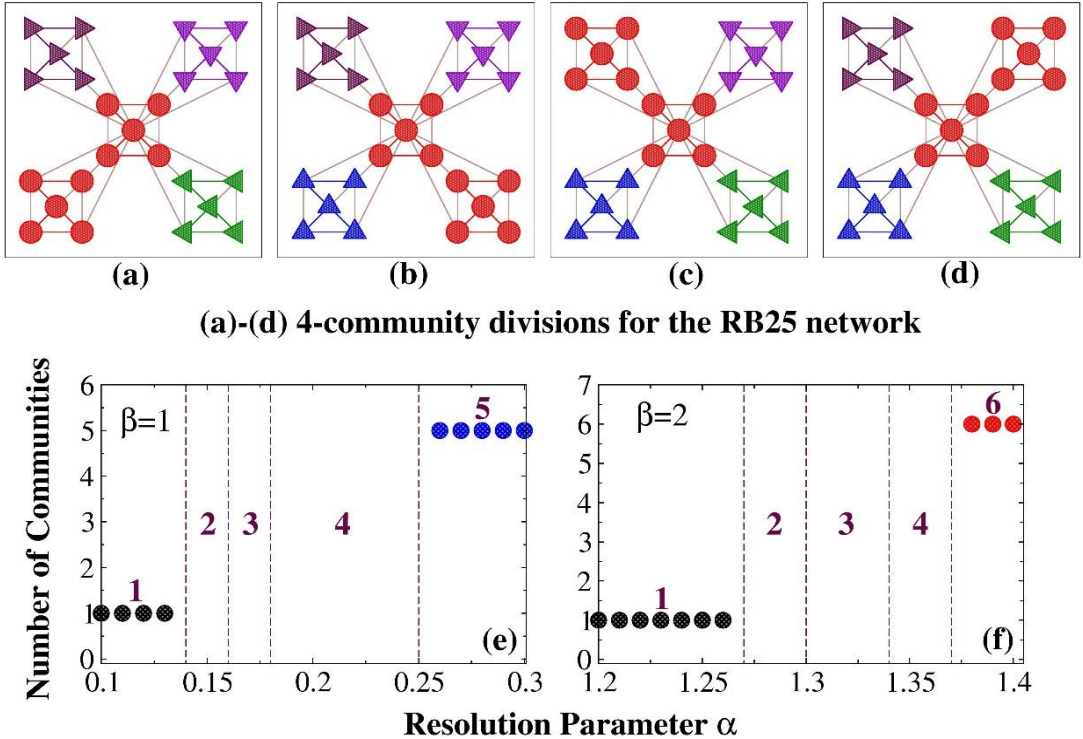

Figure 8. (color online) Unstable community structures for the RB25 network. (a)-(d): Four different 4-community divisions for the RB25 network: the central RB5 unit is *randomly* combined with one of the peripheral RB5 units. All these divisions have the same value of fitness or modularity, and in certain resolutions scales, they can all be best solutions. But in the context that nodes of the network are distinguishable, these "best solutions" are non-unique thus are not qualified to contribute to our plateaus. (e)-(f): Unstable resolution scales for the RB25 network detected with $\beta=1$ and $\beta=2$. Solid circles exhibit parts of the plateaus shown in figure 1 (d) and (e), and the resolution scales between them are unstable. Within these unstable scales we only detect *best but non-unique* solutions, whose numbers of communities are tagged on the corresponding subintervals of resolution.

**Multilevel communities in real-world networks**

In previous literature, real-world networks are rarely considered as multilevel networks, even if studied by multiresolution methods. People often artificially select their favorite communities to interpret, and ignore all the others. Our study in this paper reveals that, even filtered by our strictest strategies, real-world networks still exhibit multilevel structures, whose topologies are surprisingly all best and unique. To our viewpoint, structures suggested by stable plateaus should all have their own particular information [3]: some of them have been interpreted as being relevant to the *a priori* knowledge, or metadata of the network, while the rest still awaits proper interpretations. We believe that for real-world networks, resolution scales are also essential: communities detected in different resolution scales capture different aspects of the network structure, and should be interpretable by different node attributes or metadata. For example, for the karate club network, our detection on the 2-community level captures the communities all in a strong sense, while on the 4-community level it recovers the metadata. For the dolphins' network, our 2-community level corresponds to a "known division" during the absence of individual SN100, while the 7-community level is consistent with Newman-Girvan's division, which is presumably for the period that SN100 was present. Exploring the relationship between the detected communities and the metadata is a challenging work, yet it is believed to be promising to yield insights of genuine worth [47]. Multiple resolutions apparently provide more information than fixed resolutions, thus investigating real-world networks in a multi-resolution inspection will be worth considering in future studies.

**Computational complexity of our approach**

Finally, we briefly discuss the computational complexity of our approach. It has been reported that the Louvain algorithm has linear complexity on typical and sparse data [38], which hasn't been altered in this paper since we directly used the openly accessed code provided by the authors of the algorithm. Computational complexity introduced by our approach is mainly due to the calculations in multiple realizations.

(1) For each group of fixed parameters ($\alpha$, $\beta$), we run the algorithm in 1000 realizations to make sure there are always some realizations converging to a best solution. In practice, this is usually not necessary; 100 realizations will be sufficient for most practical purposes.

(2) For some networks whose fitness landscape is really complex, involving large numbers of local maxima [33], it is sometimes difficult to filter out unstable solutions within finite realizations. Then a little trick may help to reduce the computational burden. We can run our computations in multiple batches. Each batch consists of a certain number of realizations, and produces an independent set of plateaus by the strategy proposed in section 2.3. Next we take an *intersection* over all sets of plateaus obtained in different batches: if at a certain resolution different batches yield different network divisions, this resolution will be considered irrelevant and knocked out from the plateaus in the final output. Namely, by such an intersection we are requiring not only "one plateau one topology," but also the uniqueness of this topology over multiple batches of computations. We tested on all the synthetic and real-world networks studied in this paper. By 20 batches of 100 realizations of computations, we can efficiently remove unstable results that may sometimes require almost 10000~20000 realizations in a single batch to remove them. Yet for most networks, running multiple batches is not necessary for practical purposes.

(3) For each fixed $\beta$, we vary $\alpha$ from 0 to $2\beta$-1 with a stepwise increment $\Delta\alpha$=0.01, in order to search every inches of the resolution scales and discover all potential plateaus. In practice, to reduce the computational burden, we suggest to firstly use a relatively larger value of increment $\Delta\alpha$ for a global and coarse-grained search, and then use smaller values for detailed searches in the focused regions found in the global search.

To summarize, with a classical Louvain algorithm, our method can be implemented efficiently on various classes of complex networks with acceptable computation time.

## 5. Conclusion

Based on the community fitness function firstly proposed in [20], we made two improvements. First, we introduced a scaling factor $\beta$ that amplifies the varying range of the resolution parameter $\alpha$, which also improves the scalability of the community fitness function. With this improvement, our renewed method outperforms previous methods since it not only performs excellently on large heterogeneous LFR networks without being affected by the resolution limit problem (section 3.2), but also detects multilevel communities, including the intermediate levels, in deep hierarchical RB networks (section 3.1).

The second improvement we made is that we suggested a strict definition for the term "plateau," as well as a strict strategy to identify the plateaus (section 2.3). This has on the one hand avoided the ambiguous use of the term as in previous literature, and on the other hand remarkably improved the stability of our outputs. Consequently, our method automatically removes redundant results such as randomly detected communities as in figure 8, without any artificial or arbitrary selections. Our output is very clean; all plateaus represent stable and unique divisions for the network. On synthetic networks, our plateaus all correspond to the implanted communities, without any "junks."

Applied to real-world networks (section 3.3), our method discovers multilevel communities, which are all stable and unique. Some of them correspond to known attributes, or metadata that we have about the network. At different resolutions, our detected communities capture different aspects of the network structure.

Finally, our method can be implemented with fast heuristic algorithms. In this paper, we carried it out with a classical Louvain algorithm, which turns out to be efficient in detecting communities within various types of complex networks.

Our work in this paper has revealed the advantages of one class of *scalable* community quality functions. Their outperformance on both heterogeneous and hierarchical networks notwithstanding, for community detections on extremely large or very deep networks, community quality functions with specifically developed properties (such as higher sensitivity, or the ability of zooming in certain intermediate community levels), as well as more advanced optimization algorithms, are definitely worth pursuing in future studies.

## Acknowledgement

This work was supported by the "One Thousand Talents Program" of Sichuan Province with Grant No. 17QR003, the Doctoral Research Funding with Grant Nos. 16zx7112, KZ001212 and 21zx7115, Southwest University of Science and Technology, Mianyang, Sichuan, P.R. China.

## Availability of tools and data

All tools and data used in this paper are publicly available.

Freely available code for the Louvain/BGLL algorithm can be downloaded from the webpage of Vincent Blondel: https://perso.uclouvain.be/vincent.blondel/research/louvain.html. We did not modify the source code except substituting our own community fitness function for the original modularity function.

Source code to create the LFR benchmark networks can be downloaded from: https://sites.google.com/site/santofortunato/inthepress2.

Real-world network data can be downloaded from the webpage of Mark Newman: http://www-personal.umich.edu/~mejn/netdata/. Among them, corrections to the metadata for the dolphins' social network can be obtained in Appendix C2 of reference [57].

## Supplementary material: relevant range of the resolution parameter α

In this section, we make a rough estimation on the relevant range of the resolution parameter α for each fixed scaling factor β in the community fitness function

$$F_\alpha^\beta = \sum_{\mathcal{G}} \frac{(k_{in}^{\mathcal{G}})^\beta}{(k_{in}^{\mathcal{G}} + k_{out}^{\mathcal{G}})^\alpha} \qquad \text{(Formula 3 in the main text)}$$

It has been discussed in [33] that the resolution limit problem in community detection is due to two opposite tendencies (or *biases*): the tendency of merging small communities into larger ones, and the tendency of breaking large ones into smaller pieces. These tendencies/biases can often occur simultaneously. A strict deduction for a "relevant" resolution range, within which both biases can be avoided, is not straightforward—nor is it necessary for our purpose in this paper. In this section, we investigate each bias separately, and then give a rough estimation on the bounds of the relevant range for the resolution parameter α in formula 3. It should be noted that our purpose is very simple: *all we need is a rough range for α to vary in*. Therefore, we only investigate *necessary conditions*, rather than sufficient or necessary-and-sufficient conditions.

### I. Upper bound of α: splitting a random graph

With fixed β, since large values of α deliver small communities, the upper bound of α can be estimated by the limit at which the fitness function $F_\alpha^\beta$ starts to split a graph inappropriately into smaller parts. For such an estimation, one useful reference is the random graph: since a random graph is believed to have no communities, by any algorithm it shouldn't be split into smaller pieces [33].

Suppose we have a random graph consisting of $N$ nodes, with probability $p$ each pair of nodes shares an edge between them. Consider splitting the graph into two parts: subgraph $\mathcal{M}$ contains $m$ nodes ($0 \leq m \leq N$), while subgraph $\mathcal{N}$ contains $N-m$ nodes. Both $\mathcal{M}$ and $\mathcal{N}$ are random graphs with identical connection probability $p$.

Subgraph $\mathcal{M}$ (containing $m$ nodes) is expected to have $\frac{m(m-1)}{2}p$ intra-connections, thus it has a total in-degree $k_{in}^{\mathcal{M}} = m(m-1)p$. Each node of $\mathcal{M}$ shares with each node of $\mathcal{N}$ a connection with probability $p$, thus the out-degrees $k_{out}^{\mathcal{M}} = k_{out}^{\mathcal{N}} = m(N-m)p$. Then the fitness of community $\mathcal{M}$ can be calculated as

$$\left(F_\alpha^\beta\right)^{\mathcal{M}} = \frac{[m(m-1)p]^\beta}{[m(m-1)p + m(N-m)p]^\alpha} = \frac{m^{\beta-\alpha}(m-1)^\beta p^{\beta-\alpha}}{(N-1)^\alpha} \qquad (1)$$

Similarly, the fitness of community $\mathcal{N}$ can be calculated as

$$\left(F_\alpha^\beta\right)^{\mathcal{N}} = \frac{(N-m)^{\beta-\alpha}(N-m-1)^\beta p^{\beta-\alpha}}{(N-1)^\alpha} \qquad (2)$$

Then the fitness of the whole network (with respect to a division to $\mathcal{M}$ and $\mathcal{N}$) is

$$\left(F_\alpha^\beta\right)^{\mathcal{M}} + \left(F_\alpha^\beta\right)^{\mathcal{N}} = [m^{\beta-\alpha}(m-1)^\beta + (N-m)^{\beta-\alpha}(N-m-1)^\beta] p^{\beta-\alpha}(N-1)^{-\alpha} \quad (3)$$

In comparison, if the whole network is recognized as one community (indicated by $\mathcal{M}+\mathcal{N}$), its fitness can be calculated as

$$\left(F_\alpha^\beta\right)^{\mathcal{M}+\mathcal{N}} = N^{\beta-\alpha}(N-1)^\beta p^{\beta-\alpha}(N-1)^{-\alpha} \quad (4)$$

We do not hope the random graph $\mathcal{M}+\mathcal{N}$ be split into subgraphs $\mathcal{M}$ and $\mathcal{N}$, which requires

$$\left(F_\alpha^\beta\right)^{\mathcal{M}+\mathcal{N}} > \left(F_\alpha^\beta\right)^{\mathcal{M}} + \left(F_\alpha^\beta\right)^{\mathcal{N}} \quad (5)$$

Substitute (3) and (4) into (5), we get

$$N^{\beta-\alpha}(N-1)^\beta > m^{\beta-\alpha}(m-1)^\beta + (N-m)^{\beta-\alpha}(N-m-1)^\beta, \text{ for any } 0<m<N \quad (6)$$

Inequality (6) is equivalent to the following statement, i.e., the maximum of the function

$$f_\alpha^\beta(m) = m^{\beta-\alpha}(m-1)^\beta + (N-m)^{\beta-\alpha}(N-m-1)^\beta \quad (7)$$

should be reached at $m=0$ or $m=N$.

(7) is a $(2\beta-\alpha)$-order polynomial on variable $m$; a full set of its extrema (either maxima or minima) is difficult to solve. Here we simply make a very rough estimation on the solution of (6): we notice that due to symmetry, $f_\alpha^\beta(m)$ has an extremum at $m=N/2$ (we do not care it's a maximum or a minimum). As a *necessary condition*, (6) at least requires

$$f_\alpha^\beta(0) = f_\alpha^\beta(N) > f_\alpha^\beta(\frac{N}{2}) \quad (8)$$

Substitute (7) into (8) we get

$$N^{\beta-\alpha}(N-1)^\beta > 2^{\alpha-2\beta+1} N^{\beta-\alpha}(N-2)^\beta \quad (9)$$

i.e.,

$$2^{\alpha-2\beta+1} < (1+\frac{1}{N-2})^\beta \quad (10)$$

Obviously, when $\alpha \leq 2\beta-1$, (10) can be always satisfied. In other words, when $\alpha \leq 2\beta-1$, at least a random graph would not be split into two subgraphs of the same size. This makes a necessary condition for avoiding the first bias (inappropriate splitting of large communities); in this paper, we simply take $2\beta-1$ as the upper bound of the resolution parameter $\alpha$.

**II. Lower bound of $\alpha$: merging complete graphs**

Merging small and dense communities into larger but sparser ones, reflects the resolution limit problem at the other end of the resolution scale: small values of $\alpha$ may cause this problem. Suppose we have a couple of *complete graphs*: $\mathcal{M}$ consisting of $m$ nodes and $\mathcal{N}$ consisting of $n$ nodes. If $\mathcal{M}$ and $\mathcal{N}$ are identified as two independent communities, it is straightforward to calculate their in-degrees: $k_{in}^{\mathcal{M}} = m(m-1)$, $k_{in}^{\mathcal{N}} = n(n-1)$. As for their out-degrees, for simplicity we assume $\mathcal{M}$ and $\mathcal{N}$ are *disconnected*, i.e., $k_{out}^{\mathcal{M}} = k_{out}^{\mathcal{N}} = 0$. Then the fitness of the network with $\mathcal{M}$ and $\mathcal{N}$ identified as separate communities can be calculated as

$$\left(F_\alpha^\beta\right)^{\mathcal{M}} + \left(F_\alpha^\beta\right)^{\mathcal{N}} = m^{\beta-\alpha}(m-1)^{\beta-\alpha} + n^{\beta-\alpha}(n-1)^{\beta-\alpha} \tag{11}$$

For convenience, denote $m(m-1) = a$, $n(n-1) = b$, $\beta - \alpha = k$, then (11) becomes

$$\left(F_\alpha^\beta\right)^{\mathcal{M}} + \left(F_\alpha^\beta\right)^{\mathcal{N}} = a^k + b^k \tag{12}$$

On the other hand, if $\mathcal{M}$ and $\mathcal{N}$ are merged into one large community (indicated by $\mathcal{M}+\mathcal{N}$), the in-degree, out-degree and fitness of this large community can be calculated as

$$k_{in}^{\mathcal{M}+\mathcal{N}} = m(m-1) + n(n-1) = a + b,$$

$$k_{out}^{\mathcal{M}+\mathcal{N}} = 0,$$

$$\left(F_\alpha^\beta\right)^{\mathcal{M}+\mathcal{N}} = [m(m-1) + n(n-2)]^{\beta-\alpha} = (a+b)^k \tag{13}$$

We expect $\mathcal{M}$ and $\mathcal{N}$ being identified as two independent communities, which requires

$$\left(F_\alpha^\beta\right)^{\mathcal{M}} + \left(F_\alpha^\beta\right)^{\mathcal{N}} \geq \left(F_\alpha^\beta\right)^{\mathcal{M}+\mathcal{N}} \tag{14}$$

i.e.,
$$a^k + b^k \geq (a+b)^k \tag{15}$$

Consider a function of $k$: $f(k) = a^k + b^k - (a+b)^k$, (15) is equivalent to finding out a range of $k$ within which $f(k) \geq 0$.

Since $\dfrac{df}{dk} = a^k \ln a + b^k \ln b - (a+b)^k \ln(a+b)$, without loss of generality, we assume $m \leq n$, $a \leq b$, then

$$\dfrac{df}{dk} \leq a^k \ln b + b^k \ln b - (a+b)^k \ln b = [a^k + b^k - (a+b)^k]\ln b = f(k)\ln b$$

Since $f(1)=0$, then $\left.\dfrac{df}{dk}\right|_{k=1} \leq f(1)\ln b = 0$, indicating that $f(k)$ is not increasing in the neighborhood of $k=1$. Therefore, when $k<1$, i.e., $\alpha > \beta - 1$, presumably $f(k) \geq 0$, so that (15) would be satisfied. In this paper, we take $\beta$-1 as the lower bound of $\alpha$.

Combining the above **I** and **II**, we estimate a "relevant range" for the resolution parameter $\alpha$ with a fixed value of the scaling factor $\beta$: $\beta$-1<$\alpha$<2$\beta$-1. It should be noted that this range of $\alpha$ was *roughly* estimated through *necessary conditions* rather than sufficient or necessary-and-sufficient conditions: the "real" relevant scale of resolution can be expected to fall in this range, as a proper sub-region probably—but resolution limit problem can still exist in the rest part of this range since a sufficient condition is not guaranteed here.

**Supplementary Figures**

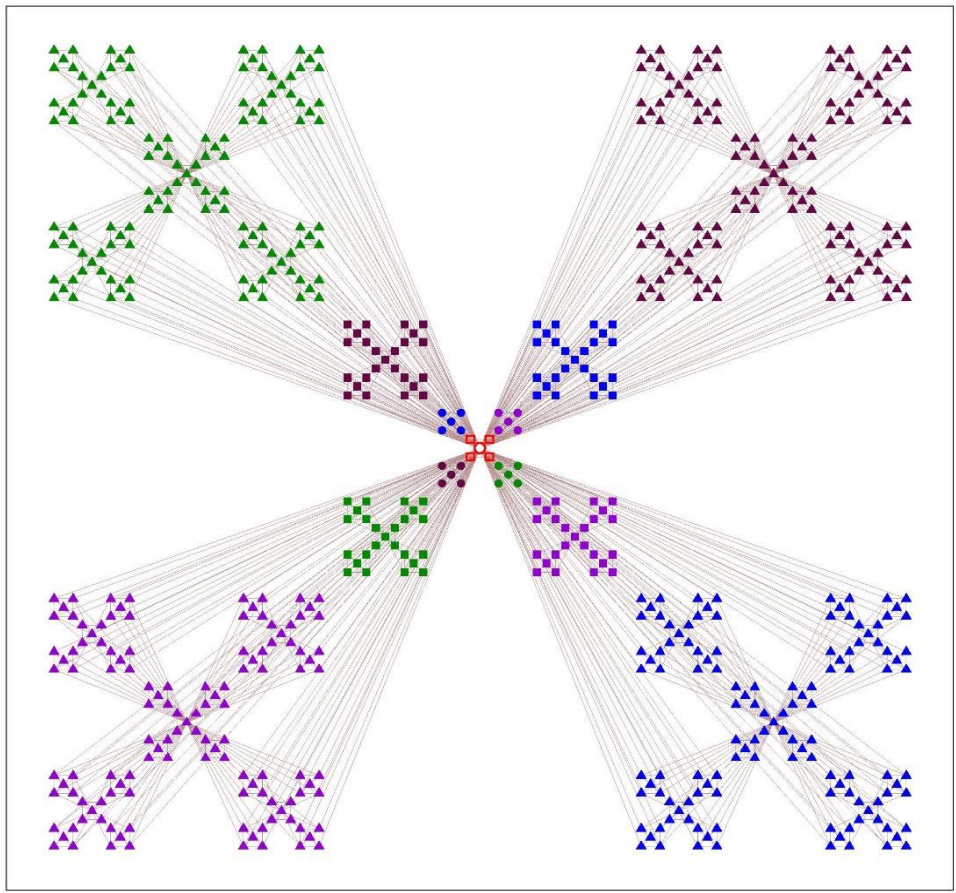

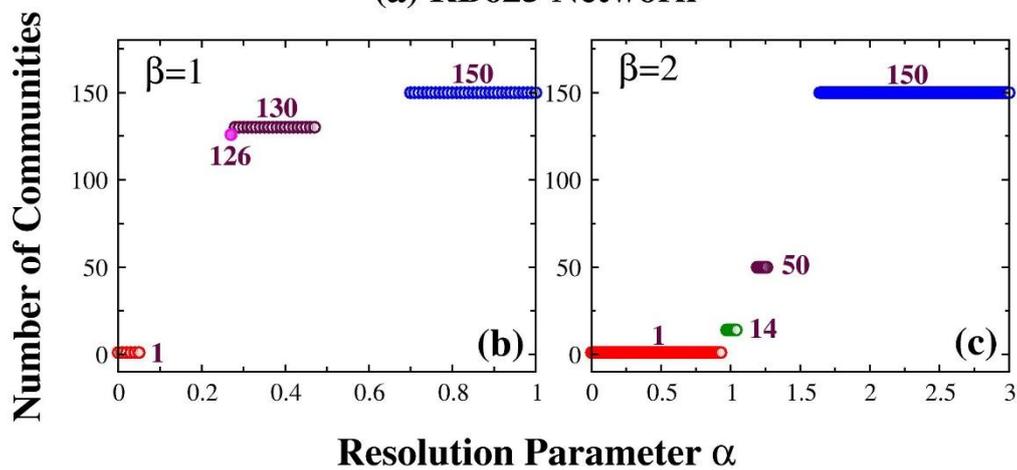

Supplementary figure 1. Communities in the RB625 network. (a) Structure of the RB625 network. Here by different shapes and colors of nodes, we show a division of the network into 14 communities on the 3rd community level. (b) and (c): Plateaus detected for the RB625 network by our method with $\beta=1$ and 2 within 1000 realizations of calculation at each resolution.

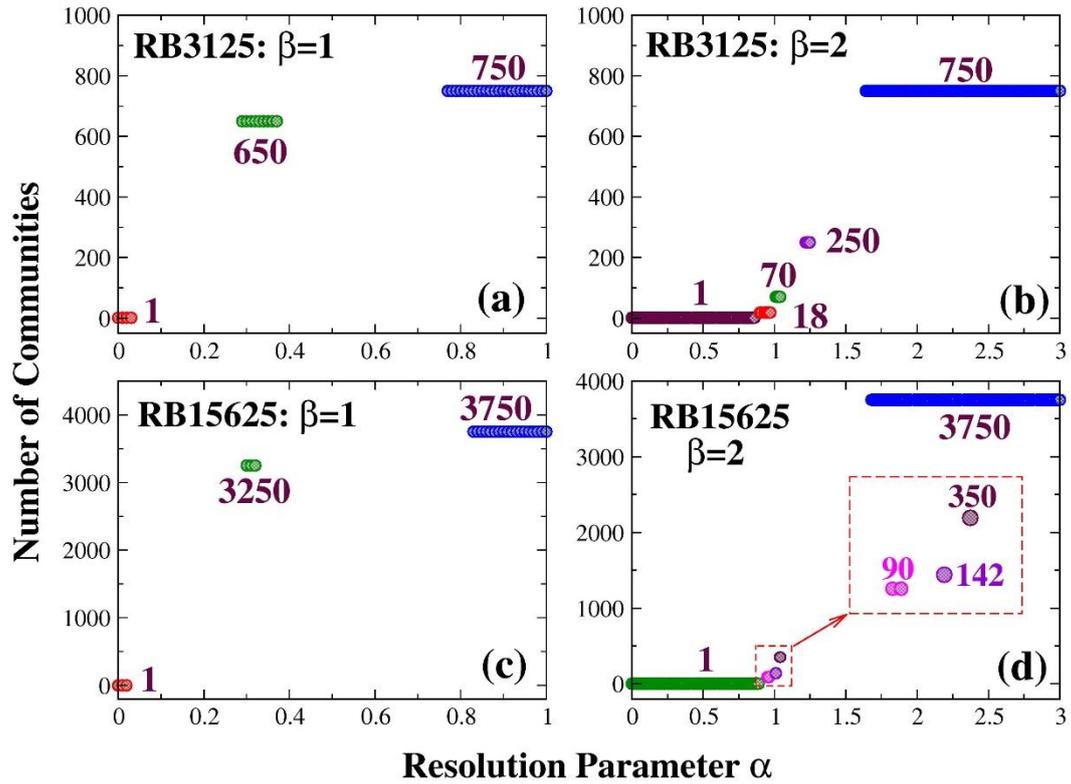

Supplementary figure 2. Plateaus detected for the RB3125 and RB15625 networks with $\beta=1$ and 2 within 1000 realizations of calculation at each resolution. Subfigures (a)-(c) are perfectly consistent with our discussion in the main text. Yet in (d), for the RB15625 network, our detection starts to deviate from perfection: (1) the 2nd and 5th community levels (containing 1250 and 22 communities respectively) are undetectable, and (2) a small plateau containing 142 communities emerges, which turns out to be a hybrid of different levels of communities: each of the four peripheral RB3125 units is divided into 18 communities (as on the 4th community level of an RB3125 network), while the central RB3125 unit is divided into 70 communities (as on the 3rd community level thereof).

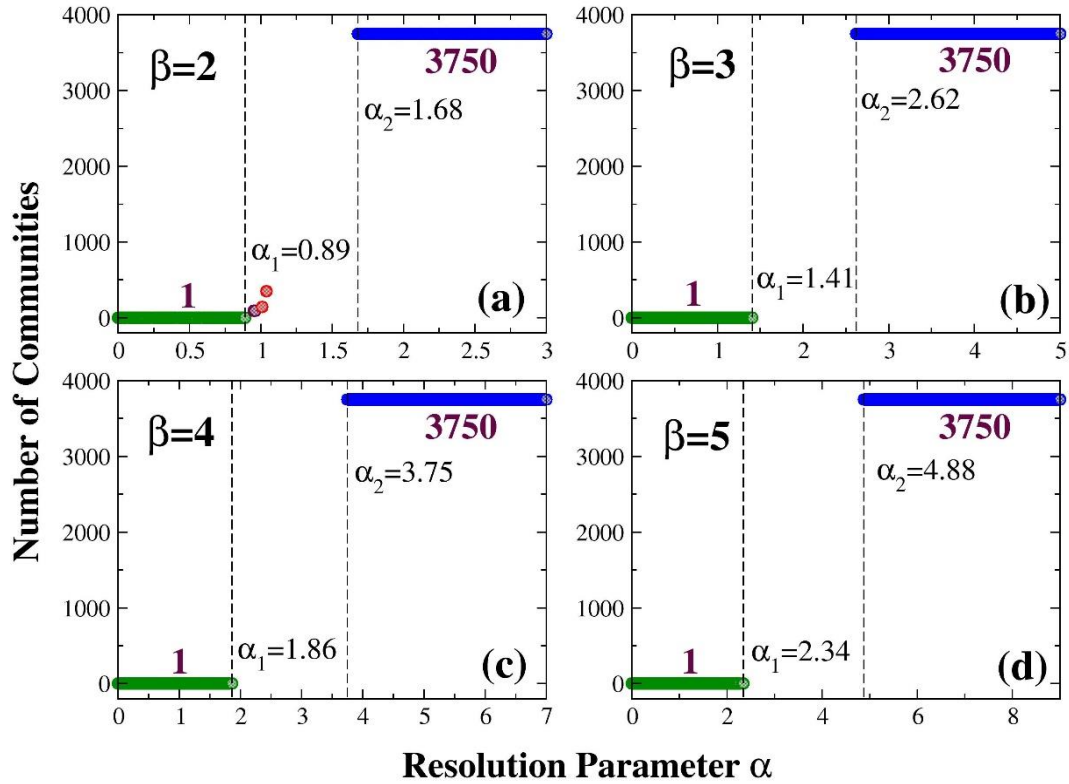

Supplementary figure 3. Plateaus for the RB15625 network detected by $\beta=2\sim5$ in 1000 realizations of calculation at each resolution. $\alpha_1$ and $\alpha_2$ in the figure indicate the upper and lower bounds of the resolution scales of the highest and lowest community levels; all intermediate community levels are limited within the range $\alpha_1<\alpha<\alpha_2$ (plateaus not shown). It can be observed that this range does not increase linearly with $\beta$; increasing $\beta$ does not help to detect more than 5 levels of communities for the RB networks.

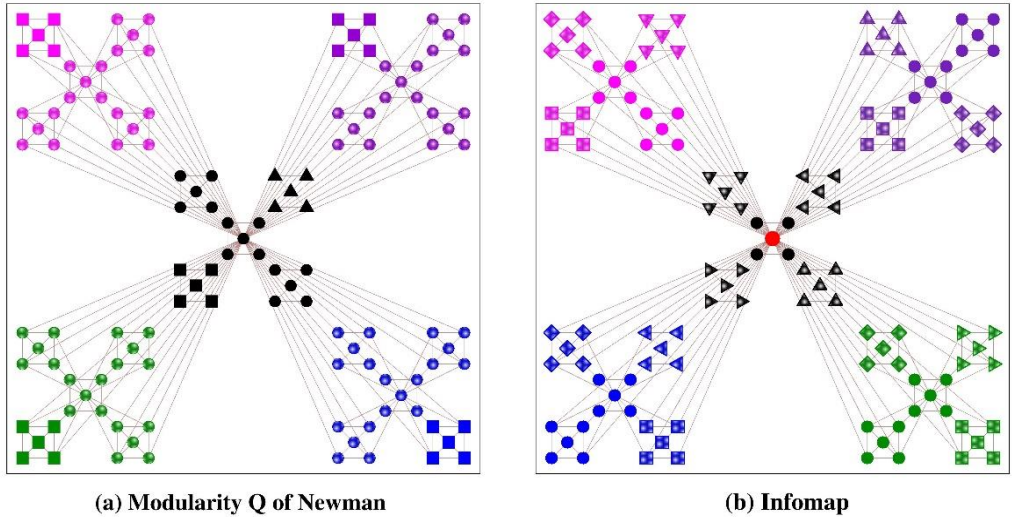

**(a) Modularity Q of Newman**  **(b) Infomap**

Supplementary figure 4. Communities in an RB125 network detected by (a) the standard modularity function $Q$ proposed by Newman [39] and (b) Infomap [21]. (a) Modularity $Q$ divides the network into 11 communities *in a random manner*: two RB5 units (black squares and triangles in (a)) are *randomly* chosen as two individual communities from the central RB25 unit, and the rest three RB5 units are integrated as a third community (black circles in (a)). Similarly, each peripheral RB25 unit is divided into two communities by *randomly* choosing one of its five RB5 units as an individual community but integrating the rest four as another community. (b) Infomap divides the network into 22 communities: the central RB25 unit is divided into six communities as in figure 1 (c) in the main text, while each of the peripheral RB25 units is divided into four communities by *randomly* combing one of its four peripheral RB5 units with the central unit, and having the rest three as three individual communities. Since both these divisions emerge with randomness, due to the symmetry of the RB networks, neither of these divisions is best-and-unique, nor are they stable solutions for community structures within the RB networks.

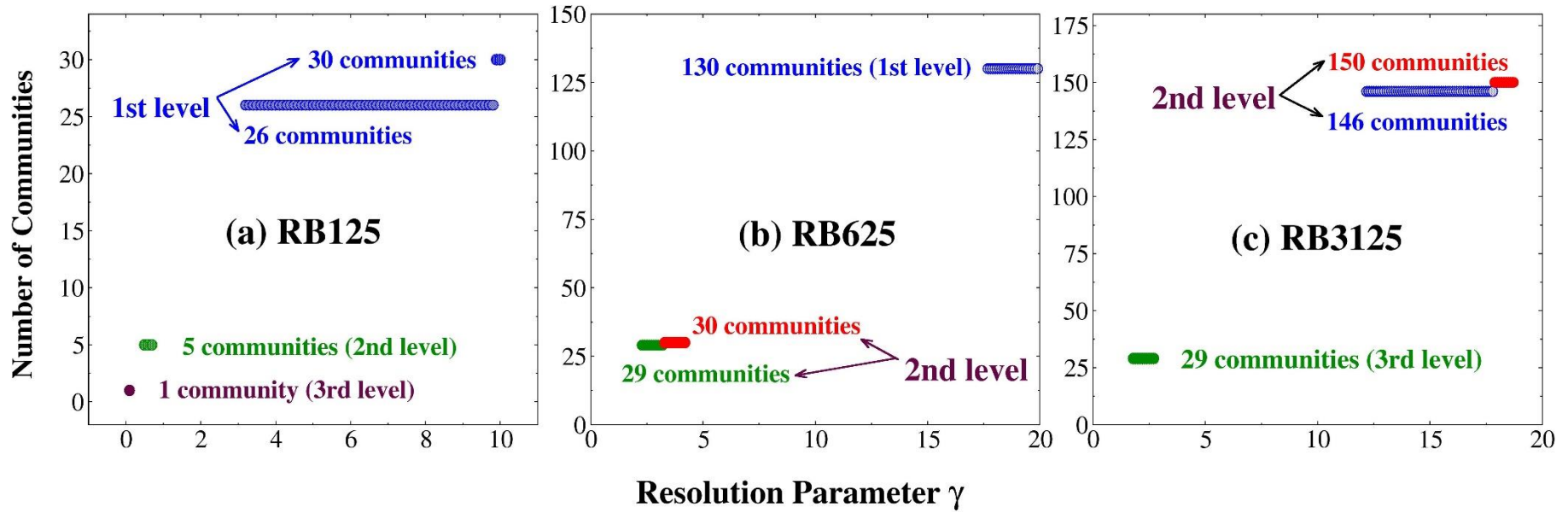

Supplementary figure 5. Plateaus identified by a generalized version of modularity $Q$ proposed by Reichardt and Bornholdt [24], i.e., $Q_\gamma = \sum_\mathcal{G} \left[ \frac{k_{in}^\mathcal{G}}{2m} - \gamma \left( \frac{k^\mathcal{G}}{2m} \right)^2 \right]$. Here the plateaus are defined with the stringency as we suggested in section 2.3, which requires "one plateau one topology" and each data point representing a "best-and-unique" solution. With such a stringency, (a) for an RB125 network, $Q_\gamma$ detects all three community levels: on the first (lowest) level, it detects both the "robust" division (30 communities) and a variant division (26 communities) which we have both discussed in section 3.1 of the main text, while on the second level, it detects only the "natural" division. (b) and (c): For larger RB networks such as RB625 and RB3125, $Q_\gamma$ detects no more than two community levels for each of them; other levels are all undetectable. Note that in (c), we have extended our calculation until parameter $\gamma$ is as large as 60, but still cannot detect a stable division of the first community level with $Q_\gamma$.

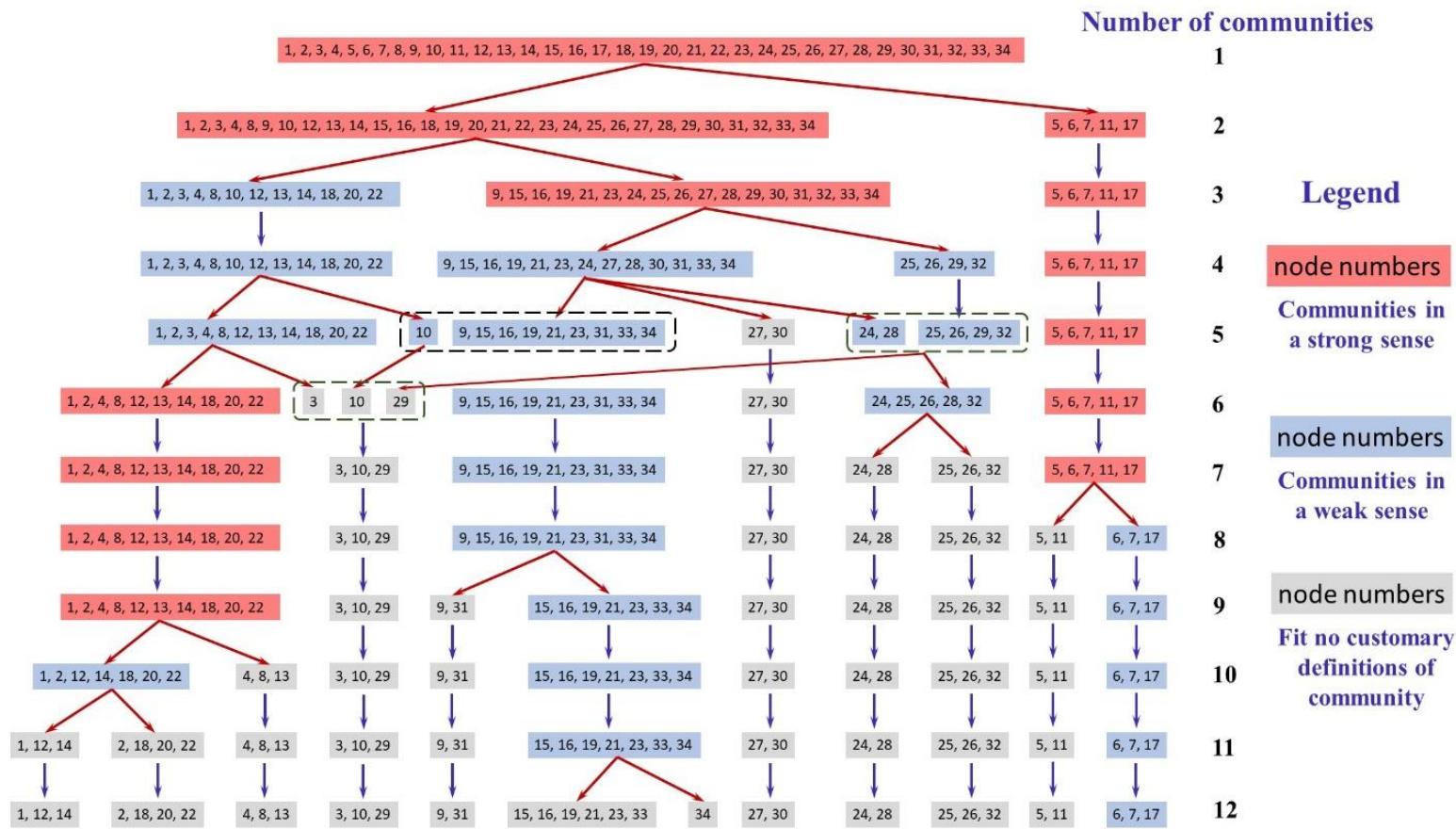

Supplementary figure 6. Multi-level community structures within the karate club network detected by our method. Different levels of communities roughly exhibit a hierarchical structure, with minor reassembling of communities (indicated by the dashed rectangles) between a few of the neighboring levels. We distinguish communities defined with different stringencies by different fill colors: red indicates communities defined in a strong sense, blue in a weak sense, while grey fits no customary definitions of community.

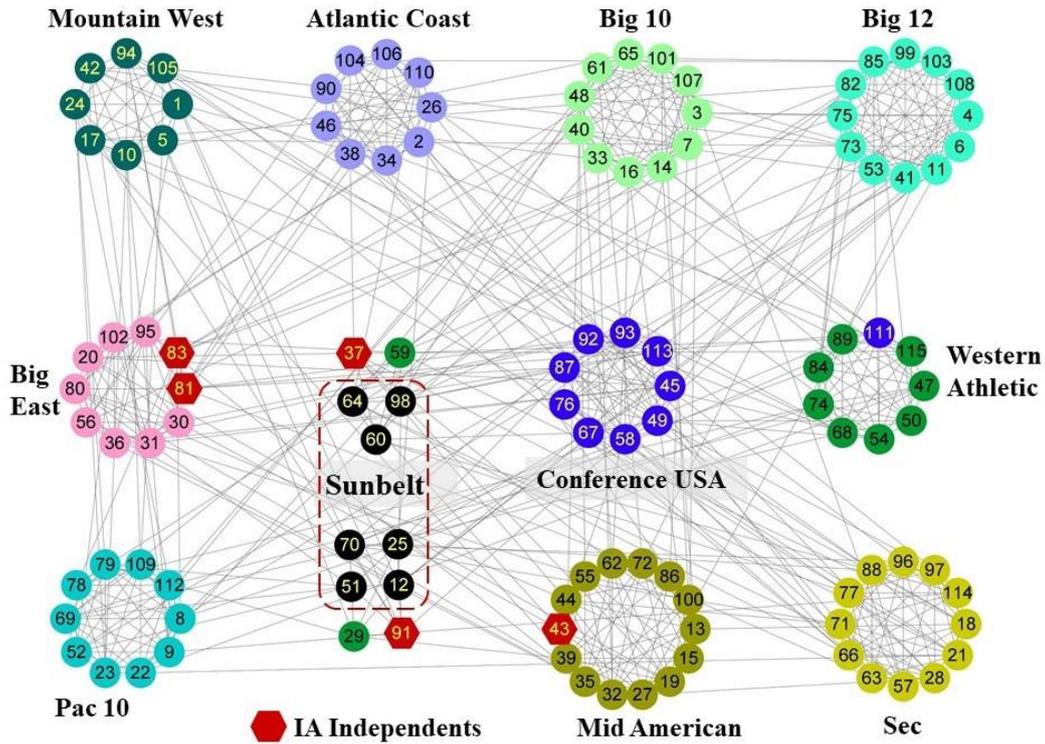

Supplementary figure 7. Twelve communities detected in the American college football network by our method, in comparison with the metadata on conference assignments recorded in figure 5 of [2]. There are some observable discrepancies between our detection and the metadata: (1) Conference Sunbelt (black nodes enclosed in the red dashed box) is split into two parts, whose members seldom played any games with the members of the other part. (2) Node 111 (Texas Christian) was recorded as a member of "Conference USA," but it did not play even one single game with any other teams of the same conference—instead it played quite some games with teams in conference Western Athletic and is then assigned to the community of the latter. (3) For the same reason, node 29 (Boise State) is assigned to one of the communities of Sunbelt instead of Western Athletic. Girvan and Newman's division agrees with ours on all the above (1)–(3); the only difference still lies in the assignment for node 37 (Central Florida). In 2010, Evans pointed out there was a serious error in the above metadata [57]: the conference assignments were collected during the 2001 season, not the 2000 season! The proof is, conference Big West existed for football till 2000 while conference Sun Belt was only started in 2001. With the metadata corrected in [57], both our division and Girvan-Newman's perfectly recreate members of all conferences (see figure 7 in the main text). As argued by the authors of [47], human errors can render the metadata irrelevant to the network structure.